\newtheorem{definition}{Definition}

\documentclass[12pt]{iopart}
\usepackage[utf8]{inputenc} % usually not needed (loaded by default)
\usepackage[T1]{fontenc}

\expandafter\let\csname equation*\endcsname\relax

\expandafter\let\csname endequation*\endcsname\relax
\usepackage[]{verbatim}
\usepackage{amsmath}
\usepackage{amsfonts}
\usepackage{mathabx}

\usepackage{graphicx}
\usepackage{subfig}
\usepackage{hyperref}
\usepackage{xcolor}

\begin{document}

\title{Percolation on the gene regulatory network}

\author{Giuseppe Torrisi, Reimer K\"uhn, and Alessia Annibale}

\address{King's College London, UK}
\ead{giuseppe.torrisi@kcl.ac.uk}
\vspace{10pt}
\begin{indented}
\item[]\today
\end{indented} 

\begin{abstract}
\noindent
We consider a simplified model for gene regulation, where gene expression is regulated by transcription factors (TFs), which are single proteins or protein complexes. Proteins 
are in turn synthesised from expressed genes, creating a feedback loop of regulation. 
This leads to a directed bipartite network in which a link from a gene to a TF exists if the gene codes for a protein contributing to  the TF, and a link from a TF to a gene exists if the TF regulates the expression of the gene. 
Both genes and TFs are modelled as binary variables, which indicate, respectively, whether a gene is expressed or not, and a TF is synthesised or not.
We consider the scenario where for a TF to be synthesised, all of its contributing genes must be expressed. This results in an ``AND'' gate 
logic for the dynamics of TFs. 
By adapting percolation theory to directed bipartite graphs, evolving according to the AND logic dynamics, we are able to determine the 
necessary conditions, in the network parameter space, under which bipartite networks can support a multiplicity of stable gene expression patterns, under noisy conditions, as required in stable cell types. 

In particular, the analysis reveals the possibility of a bi-stability 
region, where the extensive percolating cluster is or is not resilient to 
perturbations. This is remarkably different from the transition 
observed in standard percolation theory.  
Finally, we consider perturbations involving single node removal that mimic gene knockout experiments.
Results reveal the strong dependence of the gene knockout cascade on the logic implemented in the underlying network dynamics, 
highlighting in particular that avalanche sizes cannot be easily 
related to gene-gene interaction networks. 

\end{abstract}

%
% Uncomment for keywords
%\vspace{2pc}
%\noindent{\it Keywords}: XXXXXX, YYYYYYYY, ZZZZZZZZZ
%
% Uncomment for Submitted to journal title message
%\submitto{\JPA}
%
% Uncomment if a separate title page is required
%\maketitle
% 
% For two-column output uncomment the next line and choose [10pt] rather than [12pt] in the \documentclass declaration
%\ioptwocol
%
\section{Introduction}
The attempt to understand the complexity of life in terms of first principles represents a fascinating challenge of our times. The biochemical basis of life 
relies on how information is encoded in the genome and how it is expressed in living cells. 

Gene regulatory networks (GRNs) are a popular way to conceptualize the basic mechanisms involved in encoding and expression of genes at a `mesoscopic' level, without having to consider the full underlying microscopic biochemical detail. 
In this framework, genes are represented as the  nodes of a network, where edges aim to incorporate different biochemical processes linked to regulation. A state variable is associated with each node of the network, which describes the expression level of the gene.
GRNs provide useful models which
could potentially help understand biological pathways and reprogramming experiments \cite{takahashi2006induction}, and be used for the design of new drug targets \cite{10.1093/nar/gkx1314}. The rapid development of experimental techniques in the last decades has provided an enormous amount of data at the genome level \cite{grada2013next}.  Several approaches aim to infer GRNs either through experimental measurement of regulation processes \cite{zhang2008model} or through computational methods aiming at 
inferring GRNs from node states, \textit{i.e.} from gene expression  \cite{sanguinetti2019gene}.  
A popular way to study regulatory processes is to perform gene knockout experiments, which consist of the removal of nodes from the GRN \cite{zhang2014improving}. Gene expression patterns of the perturbed and unperturbed networks are compared and used to inform on the regulatory processes and derive effective gene perturbation networks
\cite{kemmeren2014large}.
In order to fully capture all collective effects in gene regulation, models used to interpret  the outcome of  these experimental protocols  would have to take  many-body interactions into account.

Kauffman's pioneering model of random Boolean networks (RBNs) accounts for  cooperative regulation in GRNs \cite{Kauffman1969437}.  In RBNs, gene states are modelled by Boolean variables, which are updated at 
regular time intervals according to a random Boolean function of the states of their neighbouring genes. It is one of the characteristic features of  RBNs to exhibit attractors in the gene expression dynamics, 
which are interpreted as stable cell types. 
RBNs aim to capture universal, \textit{i.e.} statistical, properties of 
the dynamics of synthetic random networks that are compatible with those observed in biological systems 
and have been shown to reproduce statistical features of gene 
knockout experiments \cite{serra2004149}.
While models of small Boolean networks constructed through known regulatory circuits have provided several biological insights \cite{davidich2008boolean}, 
%\textit{e.g.} the gene activation in cell cycling , 
large scale RBN lack biological interpretation and 
are hard to calibrate. Having been formulated long before the advent of the genomics era, they dot not incorporate much of the biological information which has become available in recent years.  

Nowadays, we can leverage on our improved understanding of the biological processes underpinning gene regulation, 
to make a more informed choice of the mathematical model. 
Gene regulation consists of several distinct biological processes that 
regulate gene expression. Gene expression starts with the transcription 
of a gene DNA sequence into RNA. This is generally followed by translation, i.e. the synthesis of a specific protein from the associated RNA sequence, 
although other regulatory processes are involved
\cite{schwanhausser2011global, maier2009correlation}. 
The transcription of genetic information from DNA to RNA is controlled 
by transcription factors (TFs), proteins that selectively bind to specific locations of the DNA sequence, called promoter regions, and modulate the 
expression of specific genes. TFs are themselves synthesised from expressed genes, via transcription and translation. 
The regulatory 
role of TFs in gene expression has been little investigated 
in the context of GRNs and RBN models \cite{sapienza2017dynamical}.
The regulatory effect of an individual TF binding to a given site of the promoter region of a gene is not fully known, let alone the cooperative effects of several TFs binding to several sites of the promoter region \cite{morgunova2017structural}.   
In addition, 
cooperation of TFs has been observed to play an important role to describe their binding affinity to the genome \cite{ansari2011partner}.

\begin{figure}
\centering
\includegraphics[width=0.9\textwidth]{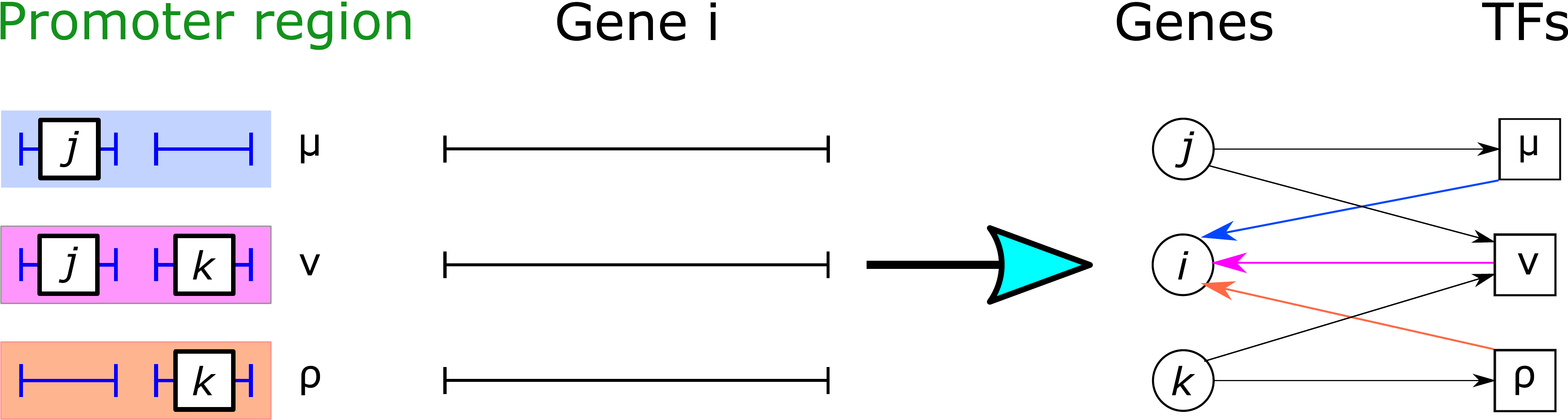}
\caption{\small Left: Diagram of the regulatory processes incorporated into the model. In this example, $\mu$ and $\rho$ are two single-protein TFs synthesised from gene $j$ and gene $k$, respectively. The TF $\nu$ is a TF complex, whose synthesis requires the simultaneous 
expression of 
$j$ {\it and} $k$. Each of the TFs $\mu$, $\rho$, and $\nu$ regulate the expression of gene $i$. Right: the same process is written in terms of a bipartite directed graph, with two sets of nodes, genes and TFs, which can be either single proteins or protein complexes.}
\label{fig:promoters}
\end{figure}

In this work, we consider a 
random Boolean bipartite network 
model
which aims to incorporate some key features of 
the biological processes summarised above, in particular, the role of 
TFs in gene expression and cooperativity effects. 
We assume that different (single-protein) TFs bind to
different sites of 
the promoter region of a gene, see Fig.\,\ref{fig:promoters} for an illustration. 
Moreover,  (single-protein) TFs can cooperatively bind to the promoter region with a potentially new regulatory power. Thus, these combinations of single-protein TFs  can be regarded as novel TFs, which are `complexes' of single-protein TFs\footnote{The term `complex' is used here and elsewhere in a loose way, to mean a combination of single-protein TFs that act together: the model is agnostic of whether they form a physical complex or not.} (see left panel of Fig.\,\ref{fig:promoters}). 
This model can be translated into a bipartite network, with two sets of nodes, i.e. genes and TFs, where TFs can be either single proteins or protein complexes. There is a directed edge from a TF to a gene if the 
former regulates the expression of the latter. Conversely, there is 
a directed edge from a gene to a TF if the former codes for a protein 
that is a constituent of the latter (see the right panel in Fig.\,\ref{fig:promoters}). %
The number of different TFs that can regulate a gene is characterised probabilistically so that the number of promoter sites associated with each gene and other microscopic details can be left open. Both genes and TFs are modelled as binary variables, which indicate, respectively, whether a gene is expressed or not, and whether a TF is synthesised or not.

A key feature of the model is that a
TF complex can only exist if {\it all} of the genes coding for proteins that constitute it are expressed {\it at the same time}. This 
fundamental level of cooperativity 
is exemplified by the choice of a specific Boolean function,  
to model the dynamics of TFs synthesis, namely an 
AND logic gate of the input gene expression levels. 
Following Kauffman's reasoning, we assume that in a given organism, all the cells have the same bipartite GRN. Hence, different cells are interpreted as distinct stable attractors of the gene expression dynamics running on the bipartite GRN, which is, in our model, 
{\it coupled} to the dynamics of TFs synthesis.
Given that all finite systems are known to be ergodic, the existence of a giant component represents the minimum requirement to support a multiplicity of stable attractors under noisy conditions, which are 
typical of biological systems. Therefore, it is crucial to know under 
which conditions, on the statistics of mutual connectivities between genes and TFs, such giant component exist. 

The existence of a giant component has been extensively studied in network science \cite{newman2001random} and its resilience against perturbations 
has been analysed in the context of percolation theory 
\cite{cohen2000resilience}. 
In this work, we investigate the percolation problem for ensembles of 
directed bipartite networks, evolving according to cooperative dynamics.
This requires developing a formalism 
which can be seen as a generalization of the homogeneous k-core percolation problem introduced in \cite{goltsev2006k}, to the heterogeneous case. 
We have recently addressed this problem in 
\cite{hannam2019percolation}, 
for Erd\"os–Rényi bipartite graph ensembles.
However, for reasons we specify below, the use of an Erd\"os–Rényi model is not fully compatible with the regulating process we aim to model. This has prompted us to re-investigate the problem concerning the existence of a giant cluster, in the context of a more suitable model class, 
which accounts for specific constraints posed by the regulatory functions a GRN is supposed to represent. 
Importantly, such re-investigation of the problem has uncovered the existence of an interesting bistability behaviour, that had 
not been made manifest in \cite{hannam2019percolation}.
In addition to re-examining the existence of a giant component, 
for bipartite graphs ensembles which are more grounded biologically, we also carry out a thorough study of the resilience of 
the giant component against perturbations which mimic gene-knockout experiments in synthetic networks. Our analysis thus provides a useful theoretical framework to interpret such experiments, quantifying,  in particular, their dependence 
on the underlying genetic logic gates.

The paper is organised as follows. 
In section \ref{sec:model} we introduce a simplified model for cooperative
gene regulatory dynamics and adapt existing definitions for the giant out- and strongly connected component to the cooperative case. In section \ref{sec:out}, we develop the formalism to predict the average fraction of nodes 
in the giant out-component, in different regions of the network parameters 
space. In section \ref{sec:out_stability}, we study the resilience of the 
giant out-component to small perturbations. In particular, we study the percolation problem in the limit where the number of removed nodes is small. In section \ref{sec:SCC} we compute the fraction of nodes in the 
strongly connected component and we prove that it is bi-stable. Finally, a summary and discussion of results is provided in Sect.\, \ref{sec:conclusions}.
The interested reader can find the code to reproduce the results shown in this paper at the following link. \footnote{ \url{https://zenodo.org/record/3798332\#.XrKo7S-ZMUs}}

\section{Model and definitions}
\label{sec:model}
\subsection{Network topology}
\label{sec:net_topo}
We construct a directed bipartite network where genes 
and TFs are the two sets of nodes. A link from a gene to a TF exists if the protein coded by the gene contributes to (i.e. is a member of) the TF. The link from a TF to a gene represents the regulation of the gene expression via promotion or inhibition of the gene expression. From a biological perspective, the latter connections aim to describe the regulatory processes related to TFs binding to the promoter region of a gene, see Fig.\,\ref{fig:promoters}. 

\begin{figure}
\centering
\includegraphics[width=0.4\textwidth]{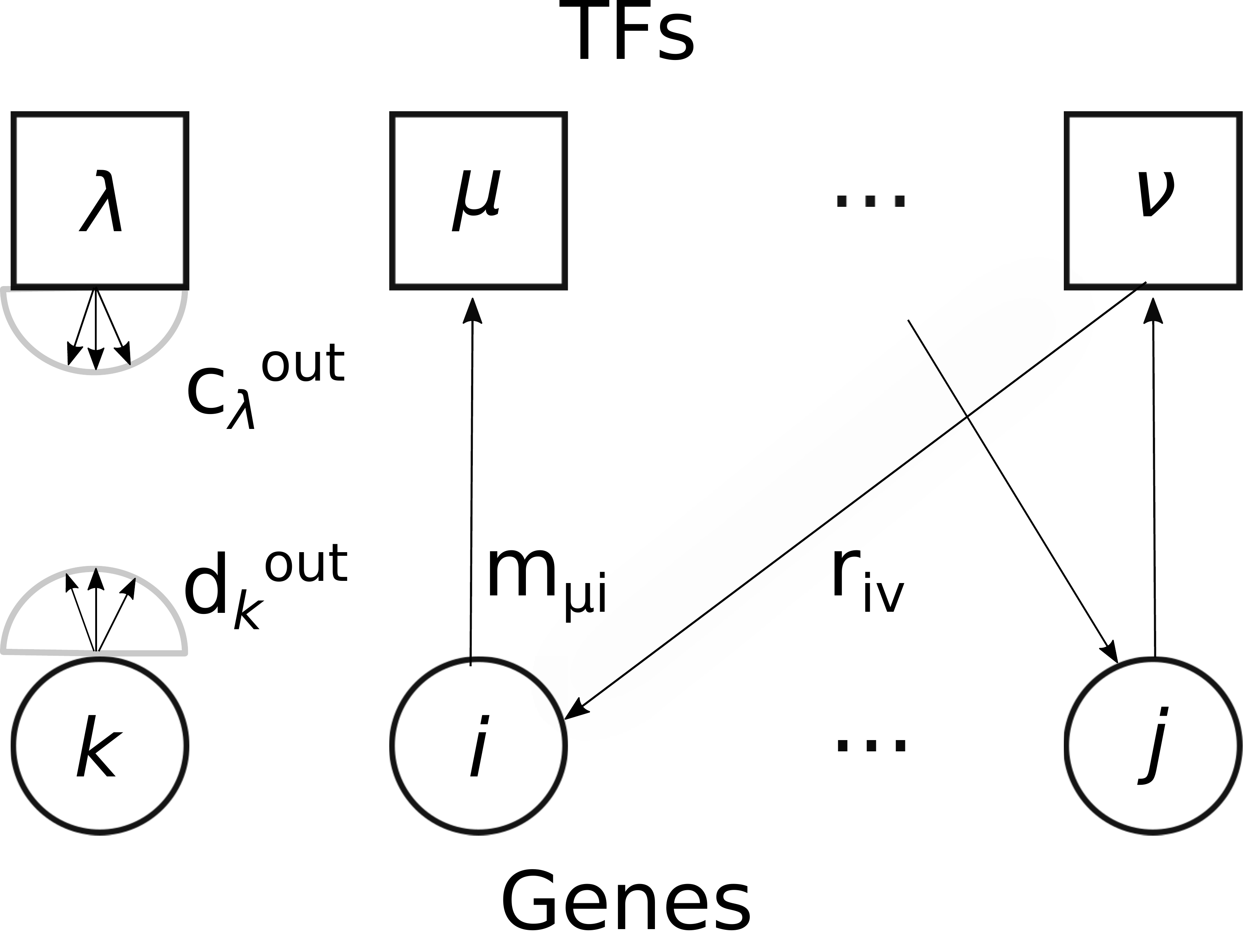}
\caption{Schematic representation of the bipartite graph. The link $m_{\mu i}$ describes the membership of gene $i$ to the production of complex $\mu$. The link $r_{i\nu}$ indicates that  TF $\nu$ regulates gene $i$.}
\label{fig:bipartite}
\end{figure}

Let  $N$ be the number of genes and $\alpha N$ the number of TFs. The bipartite graph is uniquely determined by the two bi-adjacency matrices, which are the membership matrix ${\bf m}=(m_{\mu i})$, ${\bf m} \in\mathbb{R}^{\alpha N, N}$ and the regulatory matrix ${\bf r}=(r_{i\mu})$, ${\bf r} \in\mathbb{R}^{ N,\alpha N}$, see Fig.\,\ref{fig:bipartite}. 
Their entries, regarded as random variables, 
have the following interpretation:
\begin{eqnarray}
m_{\mu i}=\begin{cases}
1, \quad \mbox{ if gene $i$ contributes to TF $\mu$}\\
0, \quad \mbox{ else}
\end{cases}
\\
r_{i\mu}\begin{cases}
>0, \quad \mbox{ if TF $\mu$  promotes transcription of gene $i$  }\\
<0, \quad \mbox{ if TF $\mu$  inhibits transcription of gene $i$  }\\
=0, \quad \mbox{ else.}
\end{cases}
\end{eqnarray}

\noindent
For any realization of the bi-adjacency matrices ${\bf r}$ and ${\bf m}$, we can define $d_i^{\mathrm{in}}({\bf r})$ as the number of TFs that 
regulate gene $i$
and $d_i^{\mathrm{out}}({\bf m})$ as the number of TFs to which 
gene $i$ contributes:

\begin{eqnarray}
d_i^{\mathrm{in}}({\bf r})&=&\sum_{\mu=1}^{\alpha N} \Theta\left( |r_{i\mu}|\right)
\nonumber\\
d_i^{\mathrm{out}}({\bf m})&=&\sum_{\mu=1}^{\alpha N}  m_{\mu i}
\label{degreeDefs_d}
\end{eqnarray}%
Analogously, we define $c_\mu^{\mathrm{in}}({\bf m})$ 
as the number of genes that contribute to TF $\mu$ (or size of the 
TF complex $\mu$)
and $c_\mu^{\mathrm{out}}({\bf r})$ as the number of genes that are 
regulated by TF $\mu$:

\begin{eqnarray}
c_\mu^{\mathrm{in}}({\bf m})&=&\sum_{i=1}^N m_{\mu i}\nonumber\\
c_\mu^{\mathrm{out}}({\bf r})&=&\sum_{i=1}^N \Theta\left( |r_{i\mu}|\right).
\label{degreeDefs_c}
\end{eqnarray}
Since promoter regions have finite length, it is plausible to assume that the sizes of TF ``complexes'' that can contribute to regulating a gene are finite, i.e. they are composed of a finite number of proteins. We also assume that TFs can only bind to a finite number of different promoter sites in the genome.\footnote{We are aware that 
this is still debated in the community,  
see \cite{kribelbauer2019low} for the TF specificity paradox.}
For these reasons, we are led to consider the case where the bi-adjacency matrices ${\bf m}=(m_{\mu i})$ and ${\bf r}=(r_{i\mu})$ are sparse. 

In the present paper, we will consider synthetic GRNs in the configuration model class, i.e. we will study ensembles of maximally random directed bipartite graphs ${\bf r,m}$ with constrained in-degree and 
out-degree sequences of genes 
$d_i^{\rm in}$, $d_i^{\rm out}$ and TFs $c_\mu^{\rm in}$, 
$c_\mu^{\rm out}$, 
drawn from the distributions
$P_{{D}}^{\mathrm{in,out}}(d)%={\rm Prob}\left(d^{\mathrm{in}}_i=d\right)
= N^{-1}\sum_i \delta_{d,d_i^{\mathrm{in,out}}}$ and $P_{{C}}^{\mathrm{in,out}}(c)%={\rm Prob}\left( c^{\mathrm{in}}_\mu=c\right) 
=\left( \alpha N\right)^{-1}\sum_\mu \delta_{c,c_\mu^{\mathrm{in,out}}}$ respectively.  We use $\langle c^{\mathrm{in}}\rangle, \langle d^{\mathrm{in}}\rangle$ to denote the mean in-degree of TFs and genes respectively, and $\langle c^{\mathrm{out}}\rangle, \langle d^{\mathrm{out}}\rangle$ to denote the mean out-degrees. 
Conservation of links implies 
$\langle d^{\rm out}\rangle=\alpha\langle c^{\rm in} \rangle$ and $\langle d^{\rm in} \rangle=\alpha\langle c^{\rm out}\rangle$. 
Without loss of generality, we will mostly set $\alpha=1$, 
corresponding to having the same number of genes and TFs.

It is important to emphasize that {\it any} 
bona fide model of a GRN must respect the following fundamental facts about gene regulation:
the regulatory part of the genome is defined by the property that genes code for at least one TF, or a component of it. Therefore, the out-degree of genes in a GRN must be larger than or equal to $1$. The same is true for the in-degree of TFs, as each TF is comprised of at least one protein. Besides, as all genes are regulated, they must, in a GRN, have an in-degree larger than or equal to $1$. However, TFs can have out-degree zero {\it in the regulatory sector} of the genome, as they may regulate other genes that are not themselves regulatory. This leads to the 
following biological constraints on the degree distributions: 
$P^{\mathrm{in}}_{D}(0)=0$, $P^{\mathrm{in}}_{C}(0)=0$, and $P^{\mathrm{out}}_{D}(0)=0$.

In the following, we will derive a general theory for {\it arbitrary} 
degree distributions, satisfying these constraints.  
Our analysis will show that different degree distributions can lead to 
different behaviours. To exemplify results, we will consider two 
families of degree distributions which display 
different behaviour. In the first family, the 
in-degrees of genes and TFs are random variables obtained as $d_i^{\mathrm{in}}=1 +\tilde d_i^{\mathrm{in}}$ and  $c_\mu^{\mathrm{in}}=1+\tilde c_\mu^{\mathrm{in}}$ respectively, where $\tilde d_i^{\mathrm{in}}$ is a random variable drawn from a Poisson distribution with average $\langle d^{\rm in} \rangle-1$ and $\tilde c_\mu^{\rm in}$ is a Poisson random variable with average $\langle c^{\rm in} \rangle-1$  
\cite{batagelj2005efficient}. The out-degrees of genes and TFs are given by $d_i^{\mathrm{out}}=1 +\tilde d_i^{\mathrm{out}}$ and $c_\mu^{\mathrm{out}}$ respectively, where $\tilde d_i^{\mathrm{out}}, c_\mu^{\rm out}$ are also drawn from Poisson distributions, with average $\langle d^{\rm out}\rangle-1$ and $\langle c^{\rm out}\rangle$, respectively.  
In what follows, we will refer to networks in this family as type I networks.
In the second family of networks (to be referred to as type II networks) the in-degrees of genes and the out-degrees of TFs are randomly sampled in the same manner as in type I networks, whereas the out-degrees of genes and the in-degrees of TFs are sampled from a discrete degree distribution $P_\gamma(k)$
with a power-law tail $P_\gamma(k)\sim \gamma k^{-\gamma-1}$ and  $\gamma>1$.  
Specifically, we will consider the 
discrete fat-tailed distribution $P_\gamma(k) = k^{-\gamma}-(k+1)^{-\gamma}$, which is defined for positive integers $1\le k\in \mathbb{N}$ and has the desired power-law behaviour for large $k$, but other choices could be made. 

\subsection{Dynamics}
We characterise the dynamic state of a cell in terms of time-dependent gene expression levels. We denote by $n_i(t) \in \lbrace 0,1 \rbrace$ for $i=1, \dots N$ the expression level of gene $i$ and by $\sigma_\mu(t)\in \lbrace 0, 1 \rbrace$ for $\mu \in 1, \dots \alpha N $ the expression level of TF $\mu$, both at time $t$. TFs are single proteins or protein complexes. A TF exists only if all the genes producing the components of the complex are expressed at the same time.  TFs can act as enhancers or inhibitors of gene activation. A gene $i$ is expressed if the overall regulatory effect of TFs that are enhancers exceeds the total regulatory effect of inhibitors by a suitable margin. Assuming a linear model for the combination of regulatory effects of the TFs contributing to the regulation of a gene, we thus have a dynamics of the form
\begin{eqnarray}
    n_i(\tau+1)&=&\Theta \left[\sum_{\mu}^{\alpha N} r_{i\mu} \sigma_\mu(\tau) -\theta_i-T z_i(\tau)\right],
    \label{eq: dynamicsg}\\
    \sigma_\mu(\tau)&=&\prod_{i\in \partial _\mu^{\textrm{in}}}n_i(\tau)\ .
    \label{eq: dynamics} 
\end{eqnarray}
Here $\Theta$ is the Heaviside theta function, for which $\Theta(x)=1$, if $x > 0$,  and $\Theta(x)=0$, if $x \leq 0$, and $\theta_i$ is the activation threshold. The set of predecessors of TF $\mu$, \textit{i.e.}, the set of proteins contributing to TF $\mu$, is denoted by  $\partial _\mu^{\textrm{in}}$, and for the present purposes we can identify proteins with the genes coding for them. Gene expression is a noisy process that we model by introducing random variables $z_i(t)$, which  we take to be  i.i.d.\,zero-mean, unit-variance random variables; the parameter $T$ is introduced to parametrise the noise level in the dynamics.
Depending on the choice of the regulatory and membership matrices 
${\bf r}$ and ${\bf m}$ and of the threshold $\theta$, the 
dynamics described by Eqs.\,\eqref{eq: dynamicsg}, \eqref{eq: dynamics} is able to capture a rich spectrum of Boolean functions in regulation \cite{buchler2003schemes}.

\subsection{The role of percolation}

The gene expression dynamics, as formulated above, must be compatible with the existence of multi-cellular organisms. Associating cell types with stable attractors of the dynamics in the noiseless limit, and assuming that different cell types share the same GRN, we thus have to require that the dynamical system 
allows for the existence of a multiplicity of stable attractors in the noiseless limit, which need to survive as stable phases, under (realistic) noisy conditions. 

In this respect, it is, however, important to note that finite stochastic systems cannot sustain a multiplicity of stable attractors 
as they are always ergodic: any configuration of the state space is reachable as the state of each node can be flipped due to noise. 
Thus, to have a multiplicity of non-trivial stable phases in the noisy dynamics, we need the set of connected nodes in the GRN to be large. The formal condition is to require it to be a finite fraction of the entire network in the limit of an infinite number of nodes. As we are in the case of the GRN dealing with a sparsely connected complex network, this requirement is equivalent to impose that the connectivity of the networks must be such as to guarantee the existence of a giant (or percolating) connected component in the GRN. Thus, as already noted in \cite{hannam2019percolation}, percolation plays a crucial role in sustaining multi-cellular life.

The existence of a giant component has been studied in several contexts in network science \cite{newman2001random, newman2018networks}. Newman applies the generating function formalism  to compute the fraction of nodes in the giant component, which is exact on graphs with no loops \cite{newman2001random}. It self-consistently determines the probability that a node belongs to a giant component in terms of the probability that it is connected to a node that we already know to belong to the giant component. The so-called ``percolation problem'' consists more generally in identifying the set of nodes that belong to the extensive connected component even after random node or bond deletion, and it 
has been used to quantify network resilience \cite{annibale2010network}. This extensive set of connected nodes is called ``percolating set''. In the case of random node or bond removal, the probability that a node belongs to the percolating set is not uniform, but it depends on the micro-structure of the network \cite{kuhn2017heterogeneous}.  
The fraction of nodes in the giant component can be determined using the cavity method \cite{shiraki2010cavity}.  The cavity method relies on the approximation that states of branches linked to a node $i$ are statistically independent of one another, once the state of  the node $i$ is known.  This approximation holds if the graph is at least locally tree-like \cite{dembo2010ising}, and become exact in the large $N$ limit.

The existence of a giant component under the logical rules imposed in the present model can be interpreted as a heterogeneous k-core percolation problem in directed networks. Homogeneous directed k-core percolation has a game theory interpretation \cite{bhawalkar2015preventing}: a node requires a cost $k$ to remain engaged, and it receives profit $1$ from in-coming engaged neighbours. Nodes remain engaged if the payoff is non-negative. In the present case, the situation is more involved, as there are two sets of nodes: genes which have cost $1$, and TFs which 
have a cost equal to their in-degree. In comparison to their in-degree, these two sets of nodes thus have the highest and lowest possible cost; it is, therefore, non-trivial to predict the configuration that will be stable given the above rules.

\begin{figure}
    \centering
        \subfloat[][]{\includegraphics[height=0.2\textheight]{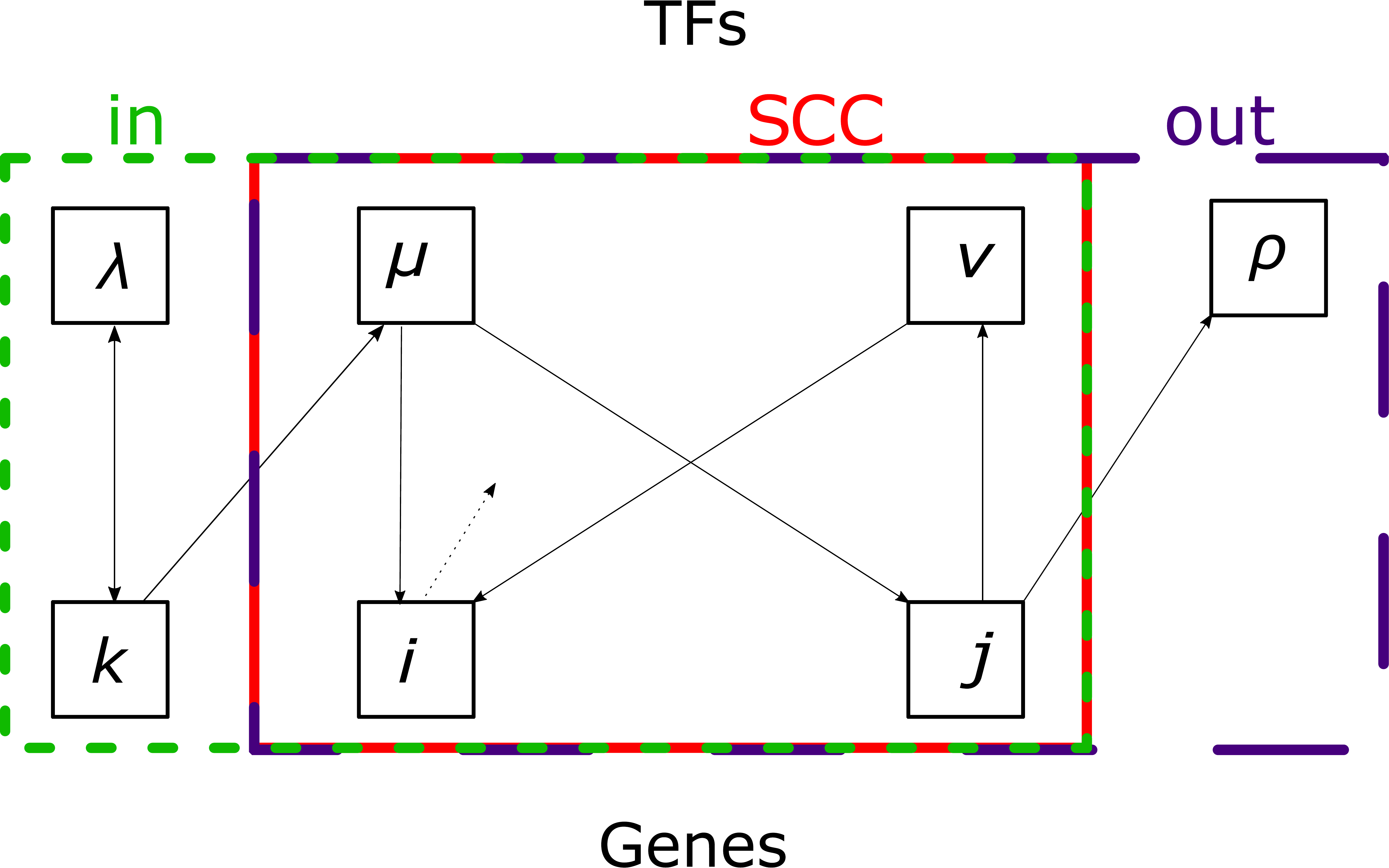}} \qquad 
        \subfloat[][]{\includegraphics[height=0.2\textheight]{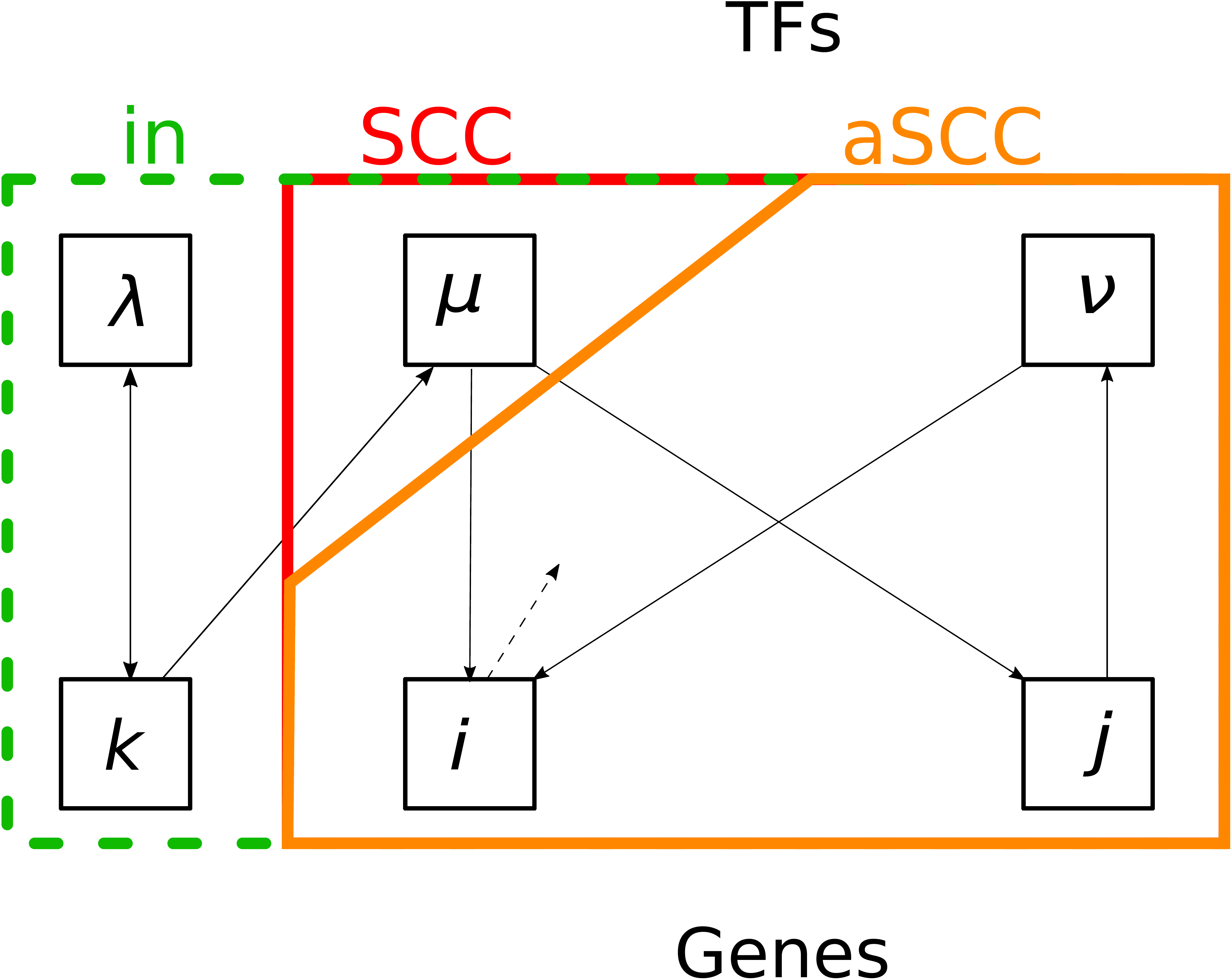}} 
    \caption{(\textbf{a}) Representation of in-component (green dotted line), ``AND'' strongly connected component (red solid line), and ``AND'' out-component (blue dashed line). (\textbf{b}) Difference between SCC and aSCC. The TF $\mu$ belongs to SCC but does not belong to aSCC.}
    \label{fig:in-SCC-out}
\end{figure}

For  (conventional) directed graphs there are several different types of giant component one can consider. They are the strongly connected component (SCC), the in-component, and the out-component, which are formally defined as follows,  see Fig.\,\ref{fig:in-SCC-out}.
\begin{definition}[Strongly connected component (SCC)]
The SCC is the set of nodes that are mutually connected by a directed path (which is itself contained in the SCC).
\end{definition}

\begin{definition}[In-component]
The in-component is the set of nodes of the graph from which the SCC is reachable via a directed path, i.e. the set of ancestors of the SCC.
\end{definition}
\begin{definition}[Out-component (OC)]
The out-component is the set of nodes of the graph that can be reached via a directed path when starting the path in the SCC, i.e. the set of descendants of a node in the SCC. 
\end{definition}

These definitions need to be adapted for the case discussed in the present paper, in which the ``AND'' logic must be imposed to ensure the existence of TFs, which in turn affects the definition of the giant components. The revised definitions of the various giant components all rest on the definition of a \textit{a valid path through a set of nodes}: a valid path through a set of nodes requires that for each TF $\mu$ on the path, all of $\mu$'s predecessors must also belong to that set. This means that for a TF to belong to a giant component, all of the genes contributing constituent proteins of the TF also need to belong to the giant component (see Fig.\,\ref{fig:in-SCC-out}) for an illustration). We thus have:
\begin{definition}[``AND'' SCC (aSCC) ]
The aSCC is a subset of a SCC  formed of nodes that are mutually connected through valid paths within the aSCC, \textit{i.e.}  a  TF $\mu$  belongs to the aSCC only if \textit{all} of $\mu$'s predecessors  belong to the aSCC as well.
\end{definition}
\begin{definition}[``AND'' out-component (aOC) ]
The aOC is a subset of an out-component formed by nodes that can be reached through valid paths originating in the aSCC.
\end{definition}
Note that the ``AND'' condition is trivially satisfied for the in-component, as for each node in the in-component, its predecessors are by definition also part of the in-component, as shown in Fig.\,\ref{fig:in-SCC-out}.
{}
In what follows, we study the existence of a giant aOC, which was previously investigated in Ref.\,\cite{hannam2019percolation}. However, 
in \cite{hannam2019percolation} a family of bipartite Erdős–Rényi graphs was considered to describe a GRN, 
which would not satisfy the constraints on the minimum in- and out-degrees that we have identified above for networks describing the gene regulatory sector. 
Re-examining the problem for a more realistic model class, 
the presence of an interesting bi-stability became manifest and revealed a shortcoming in the investigation carried out in \cite{hannam2019percolation}: there, the authors used the instability of the non-percolating solution in the equations for the aOC to define the parameter regions where a giant aOC would exist, however, this is not the right indicator in situations where the equations describing the system show bi-stability and coexistence of solutions describing percolating and non-percolating phases. 
In the rest of this manuscript, we 
address these problems, by implementing the biological constraints on the minimum degree and re-deriving the condition for the giant aOC to be stable. We obtain an analytic expression in terms of properties of the degree distribution. The derivation is performed for bipartite networks in the configuration model class which satisfy the minimum connectivity constraints and which implement the (multiplicative) ``AND" logic for the TF nodes.

\section{Existence of the giant out-component}
\label{sec:out}

We follow \cite{hannam2019percolation} to analyse conditions for the existence of a giant aOC, and determine its size in terms of both the fraction of genes and the fraction of TFs in the giant aOC.
To this end, we consider a bipartite graph of genes and TFs as described above. Indicator variables for genes and TFs are introduced, which signify whether or not they belong to the giant aOC. Thus, let $n_i=\lbrace 0, 1\rbrace$ be the indicator variable for the gene $i$,  which is 1 if $i$ belongs to the giant aOC, and 0 otherwise. 
Analogously, let  $\sigma_\mu=\lbrace 0, 1\rbrace$ be the indicator variable for TF $\mu$. Gene $i$ is in the giant aOC, if at least one of its \textit{predecessors} is also in the giant aOC (this is 
equivalent to an ``OR'' logic gate). Conversely, TF $\mu$ is in the giant aOC, if \textit{all} of its predecessors are also in the giant aOC (this is equivalent to an ``AND'' logic gate). Formally,
\begin{equation}
\begin{aligned}
   n_i & = & 1-\prod_{\mu \in \partial _i^{\mathrm{in}}}(1-\sigma_\mu^{(i)})\ ,\\
   \sigma_\mu &= & \prod_{i\in \partial _\mu^{\mathrm{in}}}n_i^{(\mu)}\ ;
\label{eq:cavity01}
\end{aligned}
\end{equation}
here $\partial _i^{\mathrm{in}}$ and  $\partial _\mu^{\mathrm{in}}$ denote the sets of TFs which are  predecessors of gene $i$ and the sets of genes that are predecessors of TF $\mu$, respectively. The variable $\sigma_\mu^{(i)}$  indicates whether the TF $\mu$ does $(\sigma_\mu^{(i)}=1)$ or does not  $(\sigma_\mu^{(i)}=0)$ belong to the giant aOC in the cavity graph from which the gene $i$ is removed; in a similar fashion, the variable $n_i^{(\mu)}$ indicates whether gene $i$ does or does not belong to the giant aOC in the cavity graph from which the TF $\mu$ is removed. 
By analogous reasoning, the indicator variables for the cavity graph must satisfy
\begin{equation}
\begin{aligned}
n_i^{(\mu)}&=&1-\prod_{\nu \in \partial _i^{\mathrm{in}}\setminus{\mu}}(1-\sigma_\nu^{(i)})\ ,\\
\sigma_\mu^{(i)}&=&\prod_{j\in \partial _\mu^{\mathrm{in}}\setminus{i}}n_j^{(\mu)}\ .
\label{eq:cavity23}
\end{aligned}
\end{equation}
Eqs.\, \eqref{eq:cavity23} constitute a set of self-consistency equations for the sets $\{n_i^{(\mu)}\}$ and $\{\sigma_\mu^{(i)}\}$ of cavity indicator variables, which can be solved iteratively for large single instances of GRNs. We are, however, mainly interested in the average fraction $g$ of genes and the fraction $t$ of TFs in the giant aOC, defined as
\begin{equation}
    g= \frac{1}{N} \sum_in_i\ ,\qquad  t = \frac{1}{\alpha N} \sum_\mu \sigma_\mu\ .
\label{eq:defgt}
\end{equation}
These can be obtained by inserting the definitions Eq.\,\eqref{eq:cavity01} into the sums in Eq.\,\eqref{eq:defgt}, and in the large system limit separately evaluating contributions coming from different in-degrees of genes and TFs respectively, giving
\begin{equation}
\begin{aligned}
g=& \sum_{d=1}P_{{D}}^{\mathrm{in}}(d)\,\left[1-(1-\tilde{t})^{d}\right]\ ,\\
t=& \sum_{c=1}P_{{C}}^{\mathrm{in}}(c)\,\tilde{g}^{c}\ .
\label{eq: gtoftgtilde}
\end{aligned}
\end{equation}
Here $\tilde t = \langle \sigma_\mu^{(i)} \rangle$ is the probability that a TF regulating a gene belongs to the giant aOC, while $\tilde g = \langle n_i^{(\mu)} \rangle$ is the probability that a constituent gene of a TF belongs to the giant aOC. In deriving Eqs.\,\eqref{eq: gtoftgtilde}, the usual approximation has been made that in large sparse, thus locally tree-like systems, averages over products of random variables involving different neighbours of a node factor. Also, it is assumed that the probability of a TF regulating a gene belonging to the giant aOC is independent of the in-degree of the gene, and in an analogous fashion, that the probability that a constituent gene of a TF is on the giant aOC is independent of the in-degree of the TF.

In a similar fashion Eqs.\,\eqref{eq:cavity23} for the cavity indicator variables can be translated into a pair of coupled self-consistency equations for the cavity probabilities $\tilde g$ and $\tilde t$. 
It will turn out, however, that, for the family of networks we are considering, 
with sparse and directed interactions and uncorrelated in- and out-degrees, the equations for the cavity probabilities 
will be identical to the equations 
for the probabilities themselves, Eqs.\, \eqref{eq: gtoftgtilde}.
%There is a subtlety, though, which entails that this pair of self-consistency equations for cavity probabilities will not take the form one might naively expect. It is related to the fact that we are dealing with sparsely connected \textit{directed} random bipartite networks. 
Due to the sparsity and directedness of the links, 
the probability that a gene is both a predecessor \textit{and} a successor of a given TF (and similarly the probability that a TF is both a predecessor \textit{and} a successor of a gene) will be inversely proportional to the system size and thus negligible in the large $N$ limit. As a consequence, the `cavity-exclusions' embodied in the product terms of Eqs.\,\eqref{eq:cavity23} will almost surely be \textit{inactive}. With these remarks in mind, we obtain
\begin{equation}
    \begin{aligned}
    \tilde{g} &=\sum_{d}\frac{d}{\langle d^{\rm out}\rangle}\,P_D^{\rm out}(d)\,\sum_{d'=1}P_D^{\rm in|out}(d'|d)\left[1-\left( 1-\tilde{t}\right)^{d'}\right]\ ,\\
    \tilde{t} &=\sum_{c}\frac{c}{\langle c^{\rm out}\rangle}\, P_C^{\rm out}(c)\sum_{c'=1}P_C^{\rm in|out}(c'|c)\,\tilde{g}^{c'}\ ,
    \end{aligned}
\end{equation}
where $P_D^{\rm in|out}(d'|d)$ is the conditional probability that 
a gene has in-degree $d'$ given that its out-degree is $d$, similarly 
$P_C^{\rm in|out}$ is the conditional probability of the in-degree of a 
TF given its out-degree. 
Here, the distributions of out-degrees of \textit{neighbours} of genes and TFs, respectively, appear due to the fact that the probability that a gene is a predecessor of a TF is actually proportional to the out-degree of the gene (and similarly for the TFs). Assuming absence of correlations between in- and out-degrees, i.e. $P_{D}^{\rm in|out}(d'|d)=P_D^{\rm in}(d')$ and 
$P_{C}^{\rm in|out}(c'|c)=P_C^{\rm in}(c')$, one finally has
\begin{equation}
\begin{aligned}
\tilde g=& \sum_{d=1}P_{{D}}^{\mathrm{in}}(d)\,\left[1-(1-\tilde{t})^{d}\right]\\
\tilde t=& \sum_{c=1}P_{{C}}^{\mathrm{in}}(c)\,\tilde{g}^{c}
\end{aligned}
\end{equation}
i.e. --- in view of Eqs.\,\eqref{eq: gtoftgtilde} --- that $g \equiv \tilde g$ and $t \equiv \tilde t$.
We thus finally have
\begin{equation}
\begin{aligned}
g=& \sum_{d=1}P_{{D}}^{\mathrm{in}}(d)\left[ 1-(1-t)^{d}\right] =1-\mathrm{G}^{\mathrm{in}}_{{D}}(1-t) &\equiv f_1(t)\ ,\\
t=& \sum_{c=1}P_{{C}}^{\mathrm{in}}(c)g^{c} = \mathrm{G}^{\mathrm{in}}_{{C}}(g) & \equiv  f_2(g)\ ,
\label{eq: out non linear}
\end{aligned}
\end{equation}
where we have also shown expressions of right hand sides in terms of the generating functions $\mathrm{G}^{\mathrm{in}}_{{D}}$ and $\mathrm{G}^{\mathrm{in}}_{{C}}$ of the in-degree distributions of genes and TFs, respectively. 

\begin{figure}[t!] 
    \subfloat[][]{\includegraphics[width=0.47\textwidth]{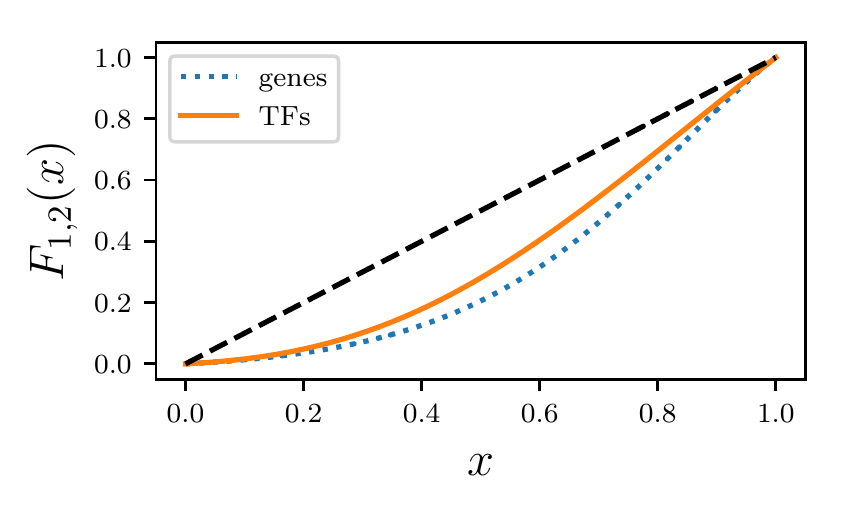}} 
      \subfloat[][]{\includegraphics[width=0.47\textwidth]{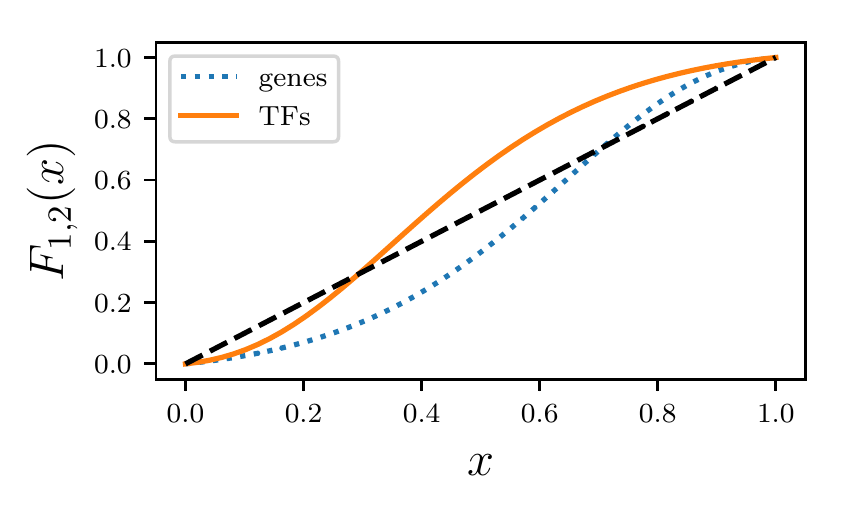}}\\
      \centering
       \subfloat[][]{\includegraphics[width=0.47\textwidth]{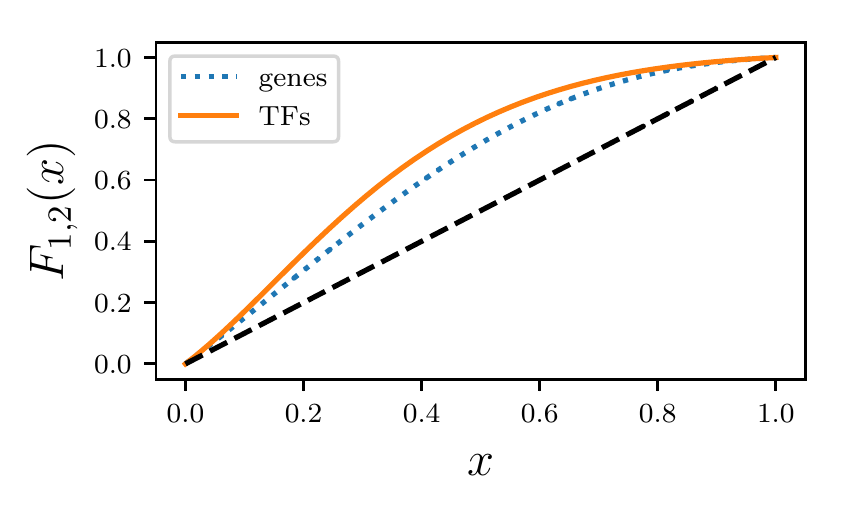}} 
\caption{\small 
Solution of Eq.\,\eqref{eq: out non linear}, obtained by intersecting the curves $F_1(x)$ (dotted line) and $F_2(x)$ (solid line), respectively, with the $x$ axis (dashed line). Results shown for Type I networks. The intersections give the fraction of genes and TFs belonging to aOC, respectively. Panels shows three different regimes (from left to right): stable solution only in $(0,0)$, bi-stability in $(0,0)$ and $(1,1)$, and stable solution only in $(1,1)$, as resulting from the stability criteria in Eqs. \, \eqref{eq:stability finite clusters}, \eqref{eq:stability_out-comp}.  
Results are shown for Type I networks with average connectivities 
$\langle c^{\mathrm{in}}\rangle=4$, $\langle d^{\mathrm{in}}\rangle=2$ (a), $\langle c^{\mathrm{in}}\rangle=4$, $\langle d^{\mathrm{in}}\rangle =4$ (b), $\langle c^{\mathrm{in}}\rangle =2$, $\langle d^{\mathrm{in}}\rangle =4$ (c). }
\label{fig:out non-linear analitic}
\end{figure}

Note that the system of equations \eqref{eq: out non linear} \textit{always} has the trivial solution $(\bar{g}, \bar{t})=(0,0)$. However, due to the  biological constraints $P_{{C}}^{\mathrm{in}}(c=0) = P_{{D}}^{\mathrm{in}}(d=0) = 0$ on the degree distributions, these equations {\it also} always have the solution $(\bar{g},\bar{t})=(1,1)$. These solutions correspond to the extreme cases where the system has no giant aOC, or the giant aOC consists of the entire graph, respectively.
A common strategy to find solutions of Eq.\,\eqref{eq: out non linear} is to obtain them as fixed points under forward iteration using the map
\begin{equation}
\begin{aligned}
g(\tau+1)=& \sum_{d=1}P_{{D}}^{\mathrm{in}}(d)\left[ 1-(1-t(\tau))^{d}\right]\\
t(\tau)=& \sum_{c=1}P_{{C}}^{\mathrm{in}}(c)g^{c}(\tau)\ .
\label{eq: out non linear_dyn}
\end{aligned}
\end{equation}
Solutions of Eqs.\,\eqref{eq: out non linear} can alternatively be obtained through a graphical method.  In Eqs. \,\eqref{eq: out non linear} one can substitute one relation into the other, to obtain the two independent equations  $g =f_1\left(f_2(g)\right) \equiv F_1(g)$ and $t=f_2\left(f_1(t)\right) \equiv F_2(t)$. In Fig.\,\ref{fig:out non-linear analitic} we show the graphs of $F_1(g)$ and $F_2(t)$ for three different choices of the connectivity in Type I networks, corresponding to (i)  a situation where the giant aOC solution is unstable  (panel a), (ii) coexistence of a solution representing a stable giant aOC and a stable solution representing the absence of a giant aOC (panel b), and (iii) existence of a stable giant aOC (panel c).

To assess the stability of the solutions  $(g,t)=(0,0)$  and  $(g,t)=(1,1)$ under forward iteration, we consider the Jacobian for the functions $f_1(t)$ and $f_2(g)$. The condition for stability under forward iteration  is that the maximum eigenvalue of the Jacobian  at a fixed point is smaller than 1, which becomes
\begin{equation}
\langle d^\mathrm{{in}}\rangle P_{{C}}^{\mathrm{in}}(c=1)<1
\label{eq:stability finite clusters}
\end{equation}
for $(g,t)=(0,0)$, and
\begin{equation}
\langle c^\mathrm{{in}}\rangle P_{{D}}^{\mathrm{in}}(d=1)<1
\label{eq:stability_out-comp}
\end{equation}
for  $(g,t)=(1,1)$.
%Note that conditions for local uniqueness of the two solutions are weaker, namely $\langle d^\mathrm{{in}}\rangle P_{{C}}^{\mathrm{in}}(c=1) \neq 1$ for $(\bar{g},\bar{t})=(0,0)$, and $\langle c^\mathrm{{in}}\rangle P_{{D}}^{\mathrm{in}}(d=1) \neq 1$ and for  $(\bar{g},\bar{t})=(1,1)$.
%
\begin{figure}
\subfloat[][]{\includegraphics[width=0.47\textwidth]{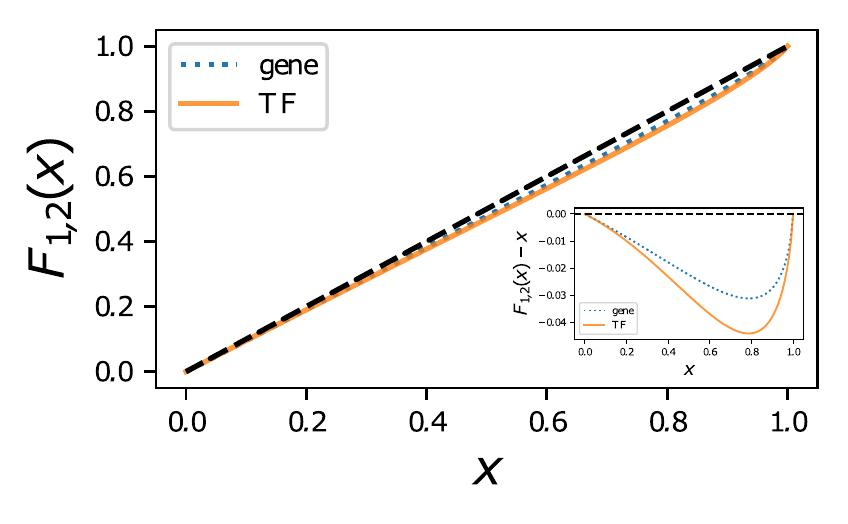}} 
\subfloat[][]{\includegraphics[width=0.47\textwidth]{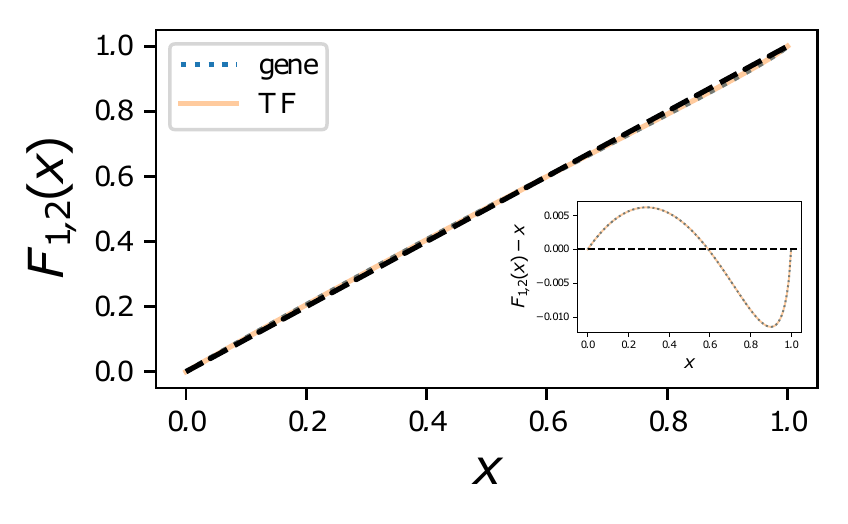}}\\
\centering
\subfloat[][]{\includegraphics[width=0.47\textwidth]{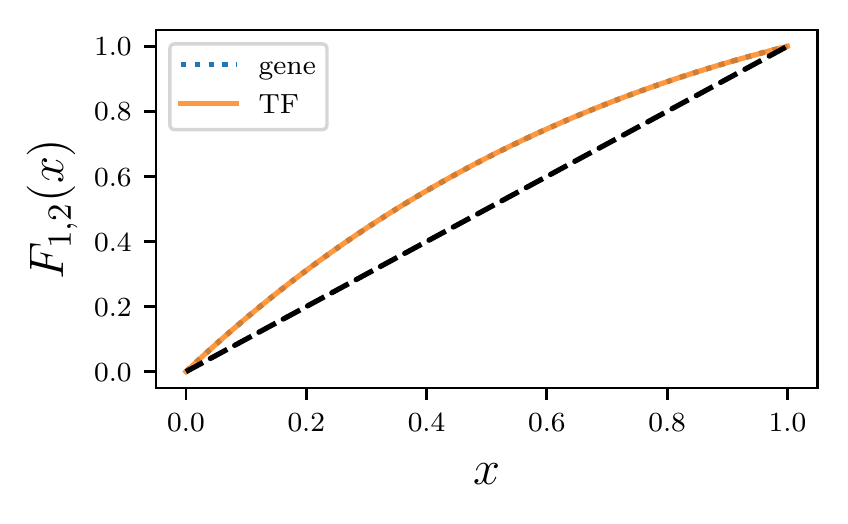}} 
\caption{Solution of Eq.\,\eqref{eq: out non linear}, obtained by intersecting the curves $F_1(x)$ (dotted line) and $F_2(x)$ (solid line), respectively, with the $x$ axis (dashed line). Same quantity as in Fig.\, \ref{fig:out non-linear analitic} for Type II networks. Panels shows three different regimes (from left to right): stable solution only in $(0,0)$, un-stable solution  in $(0,0)$ and $(1,1)$, and stable solution in $(1,1)$, as resulting from the stability criteria in Eqs. \, \eqref{eq:stability finite clusters}, \eqref{eq:stability_out-comp}. Inset in (a) and (b) shows the plot of $F_{1,2}(x)-x$, to magnify the nature of transition. Parameters of the networks: $\langle c^{\mathrm{in}}\rangle=1.8$, $\gamma=1.1$ (a), $\langle c^{\mathrm{in}}\rangle=1.8$, $\gamma =1.25$ (b), $\langle c^{\mathrm{in}}\rangle =1.8$, $\gamma =5.$ (c). }
\end{figure}
For some network distributions, 
the stability conditions of Eqs. \eqref{eq:stability finite clusters} and \eqref{eq:stability_out-comp} may be both satisfied at the same time. 
In that case, both $(0,0)$ and $(1,1)$ are stable fixed points, and the dynamical system described by the map Eq.\,\eqref{eq: out non linear_dyn}  exhibits bi-stability.  

Type I and type II networks as specified in \, \ref{sec:net_topo} are examples of networks that can and cannot realise bi-stability, respectively.  
Indeed, for networks of type I with connectivities defined in terms of shifted Poisson degree distributions, the condition Eq.\,\eqref{eq:stability_out-comp} for the stability of the $(g,t)=(1,1)$ solution can be rewritten as $\langle c^{\mathrm{in}}\rangle <\exp{\left[\langle d^{\mathrm{in}}\rangle-1\right]}$, and the complementary condition Eq.\,\eqref{eq:stability finite clusters} for the stabiltiy of the non-percolating $(g,t)=(0,0)$ solution as $\langle c^{\mathrm{in}}\rangle>1+\log\left[\langle d^{\mathrm{in}}\rangle\right]$. It is relatively straightforward to demonstrate that there is for \textit{any} choice of $\langle d^{\mathrm{in}}\rangle$ a range values of $\langle c^{\mathrm{in}}\rangle$ that satisfy \textit{both} conditions.
For type II networks on the other hand, Eq.\,\eqref{eq:stability_out-comp} becomes $\zeta(\gamma)=\langle c^{\mathrm{in}}\rangle < \exp{\left[\langle d^{\mathrm{in}} \rangle-1\right]}$, where $\zeta(\gamma)$ is the Riemann Zeta-function, and Eq.\,\eqref{eq:stability finite clusters} gives $\gamma<\log_2\left(\langle d^{\mathrm{in}}\rangle/\left( \langle d^{\mathrm{in}}\rangle-1\right)\right)$. Evaluating these conditions, 
one finds that they are never satisfied at the same time for \textit{any}  $\gamma>1$. 

 For networks that allow a range of 
connectivities where 
there is coexistence of two stable solutions, $g = 0$ and $g = 1$, a discontinuous transition 
from a percolating to a non-percolating regime 
is expected, when connectivity parameters are varied. 
Conversely, for networks where there is always only a single stable solution, as connectivity is varied, the transition from a
percolating to a non-percolating situation is expected to be 
continuous.

\section{Stability of the giant aOC and the non-percolating solution}
\label{sec:out_stability}

Equations \eqref{eq: out non linear} describe the fractions of genes and TFs that belong to the giant aOC. We have identified two solutions which postulate extreme situations, where either the entire network belongs to the aOC, or there is no giant aOC at all. 
In this section, 
we will analyse 
the stability of these two extreme situations with respect to small perturbations, which amount to either removing small amounts of genes from 
the giant aOC (in the case where it encompasses the entire system), or to `seeding' a giant aOC with a small sub-extensive aOC. 
The conditions we will obtain from the stability analysis against  perturbations will coincide with the stability conditions Eqs.\, \eqref{eq:stability finite clusters} and \eqref{eq:stability_out-comp} for the fixed points of the dynamical system Eq.\,\eqref{eq: out non linear_dyn}, 
demonstrating the equivalence of the two notions of stability.

We begin by studying the robustness of a giant aOC that encompasses the entire system, to random removal of nodes. We use the cavity method to derive the macroscopic fraction of nodes belonging to the percolating set after removal of a finite fraction of genes. In an analogous fashion, we will study the stability of the non-percolating solution by using the cavity method to obtain the fraction of nodes in an aOC which contains a small specified initial set of genes. 

In a large network, a self averaging property holds, which guarantees that the fraction of nodes in the giant aOC converges to the probability that a node is in the percolating set. We shall find that the condition given in Eq.\,\eqref{eq:stability_out-comp} provides the minimal requirement for a giant aOC to be resilient to removal of a finite number of genes, and similarly Eq.\,\eqref{eq:stability finite clusters} 
provides the minimal requirement for the size of an aOC found by ``seeding'' it from a finite set of genes, to remain sub-extensive.

However, close to the instabilities (i.e. phase transitions) predicted by the macroscopic theory, there are, as usual, pronounced finite-size effects that lead to a `smearing' of the otherwise sharp transitions. For example, if the number of nodes removed is small, i.e. it does not grow with the size of the system, simulations close to phase transitions show that the outcome of a single node deletion experiment depends on the topological properties of the nodes that are eliminated. This leads to a rounding of a transition that would be sharp in the infinite system limit.
For one such case that is close to instability, we use simulations to study the full distribution of microscopic outcomes of the percolation problem for a given network. 

Turning to the biological interpretation of the stability results, we argue that the stability of the giant aOC against the removal of a finite amount of gene can produce valuable insights for the interpretation of gene knock-out experiments, whereas the question of the local stability of the non-percolating solution is intimately related to the possibility of gene activation cascades.

\subsection{Stability of percolating phase}
\label{sec: perturbation on}
\begin{figure}
    \centering
      \subfloat[][]{\includegraphics[width=0.47\textwidth]{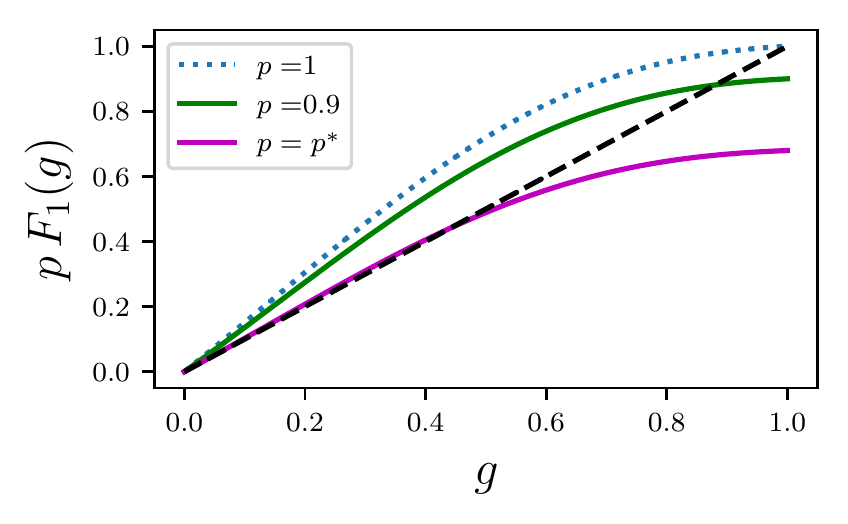}} \quad
          \subfloat[][]{\includegraphics[width=0.47\textwidth]{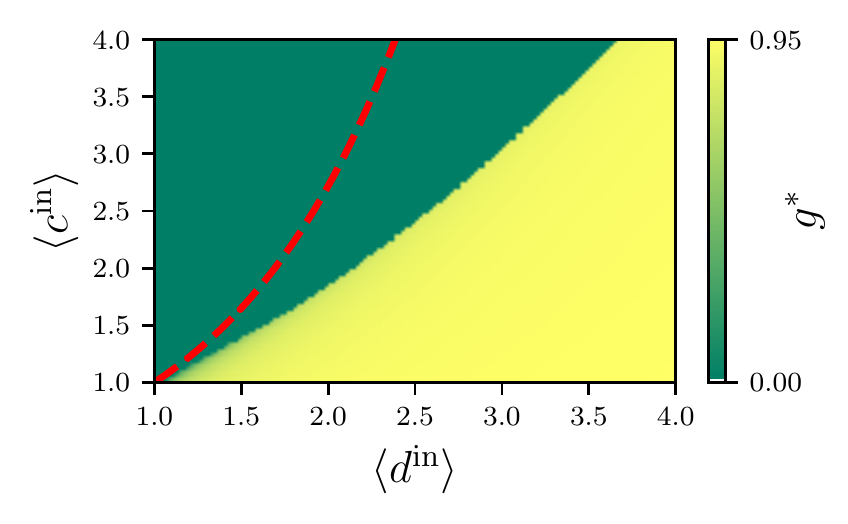}} 
   \\
    \caption{\small \textbf{(a)} Graphical solution of  Eq.\,\eqref{eq:non-linear pert}, obtained as intersection of the curve $p\, F_1(g)$ with $g$. The solid lines show the perturbed 
    cases ($p=0.9$, $p=p^*$) and the dotted line shows the unperturbed curve ($p=1$). Results are shown for type I networks with connectivity $\langle c^{\mathrm{in}}\rangle= 2$, $\langle d^{\mathrm{in}}\rangle =$4. 
    \textbf{(b)} Heatmap showing the value $g^\star$ of the largest solution of Eq.\,\eqref{eq:non-linear pert}, for different values of the mean connectivities in type I networks. The dilution parameter is $p$=0.95. The dashed line indicates the transition curve at $p=$1.}
 \label{fig:perturbed}
\end{figure}

We investigate the stability of the solution $(g,t) = (1,1)$ 
against a perturbation that deletes a certain fraction of genes from 
the network, or more precisely declares them as not being part of the original giant aOC. 
Similar perturbation protocols have been used 
recently in the context of satisfiability problems \cite{lagomarsino2005logic}\cite{correale2006core}.
Let us consider a setting where a fraction $1-p$ of genes are thus removed from the aOC. Introducing a binary random variable $\chi_i$ i.i.d., taking the value $\chi_i=1$ if gene $i$ is kept in the aOC, and $\chi_i=0$ if it is removed, we can analyse the existence of the giant aOC as a site percolation problem in terms of a set of indicator variables $n_i$ for genes and $\sigma_\mu$ for TFs, as in Sect.\, \ref{sec:out} above. Investing the almost sure equality of cavity and non-cavity indicator variables that we have argued for there, the modification of Eqs.\,\eqref{eq:cavity01} needed to capture the removal of a pre-defined set of genes is 
\begin{equation}
\begin{aligned}
n_i&=&\chi_i\left(1-\prod_{\mu \in \partial _i^{\mathrm{in}}}(1-\sigma_\mu)\right)&\\
\sigma_\mu&=&\prod_{i\in \partial _\mu^{\mathrm{in}}}n_i& \ .
\end{aligned}
\label{eq:dynamics_non-linear_perc}
\end{equation}
Averaging these over the realizations $\{\chi_i\}$ of the percolation experiment, and using the notation $g_i=\left\langle n_i\right\rangle_{\chi}$ and $t_\mu=\left\langle \sigma_\mu\right\rangle_{\chi}$ to denote averages of local indicator variables over realizations, we get
 \begin{equation}
  \begin{aligned}
 g_i&=&
p\left(1-\prod_{\mu \in \partial _i^{\mathrm{in}}}(1-t_\mu)\right)\ , \\
t_\mu&=&\prod_{i\in \partial _\mu^{\mathrm{in}}}g_i \ ,
\end{aligned}
\label{eq:cavity disorder average}
 \end{equation}
 where we have set $p=\langle \chi _i \rangle_\chi$. Taking the average over the sites of a graph, defining $g= N^{-1}\sum_i g_i$ and $t = (\alpha N)^{-1}\sum_\mu t_\mu$ as in Sect.\,\ref{sec:out}, one obtains
\begin{equation}
\begin{aligned}
g=& p\sum_{d=1}P_{{D}}^{\mathrm{in}}(d)\left[1-(1-t)^{d}\right]&\equiv &h_1(t)\\
t=& \sum_{c=1}P_{{C}}^{\mathrm{in}}(c) g^{c}&\equiv &h_2(g)\\
\end{aligned}
\label{eq:non-linear pert}
\end{equation}
The point $(g,t)=(1,1)$ is now no longer a solution. However, for certain degree distributions, there are still two stable solutions, namely 
$(g,t)= (0,0)$ and $(g,t)=(g^\star,t^\star)$. 
Let us denote the function composition 
$H(g)=h_1(h_2(g))$, such that the fraction of genes $g^{\star}$ that is a solution of \eqref{eq:non-linear pert} is given by $g^{\star}=H(g^{\star})$. The composition of functions for the perturbed setting can be written in terms of the unperturbed composition as $H(g)=p\,F_1(g)$, with $p\in [0,1]$. Hence, the curve $H(g)$ in the perturbed case, is always upper bounded by that of the unperturbed case, $F_1(g)$, see Fig.\,\ref{fig:perturbed}a. 
This implies that a network where the solution $(g,t)=(1,1)$ is 
unstable in the unperturbed case (as in Fig.\, \ref{fig:out non-linear analitic}a), will {\it no longer} exhibit a 
solution $(g,t)=(g^{\star},t^{\star})$ in the vicinity of $(1,1)$, 
when a small perturbation is applied
\textit{i.e.} Eq.\,\eqref{eq:stability_out-comp} gives the necessary condition for the stability of the giant aOC which encompasses the 
entire system. Conversely, in networks where the solution 
$(g,t)=(1,1)$ is the only stable solution at $p=1$ (as in Fig.\,\ref{fig:out non-linear analitic}c), this is expected to be 
continuously deformed when $p$ is decreased to values below $p=1$. 
The nature (continuous or discontinuous) of the phase transition 
that we have discussed for the unperturbed system in Sect.\,\ref{sec:out}
is expected to be indicative of the behaviour at $p$ below but close to $1$.  This behaviour remains qualitatively similar 
as $p$ is lowered further, away from $1$, as long as  $p > p^*$, with  $p^*=1/F'_1(0)=\left( \langle d^\mathrm{{in}}\rangle P_{{C}}^{\mathrm{in}}(c=1)\right)^{-1}$.

In Fig.\,\ref{fig:perturbed}b, we plot 
the largest solution $g^{\star}$ of Eq. \,(\ref{eq:non-linear pert}),
as a function of the mean connectivities, for bipartite graphs of 
type I. As expected, a discontinuous transition of $g^{\star}$ is exhibited as connectivities are changed.
Results are shown for $p=0.95$, and the location of the transition 
curve at $p=1$ (predicted by Eq. \,\eqref{eq:stability_out-comp}) is 
shown as a dashed line. This indicates that the condition for the emergence of a non-zero fixed point in the perturbed case is more  stringent than for the unperturbed case.

\begin{figure}
\centering
    \subfloat[][]{\includegraphics[width=0.47\textwidth]{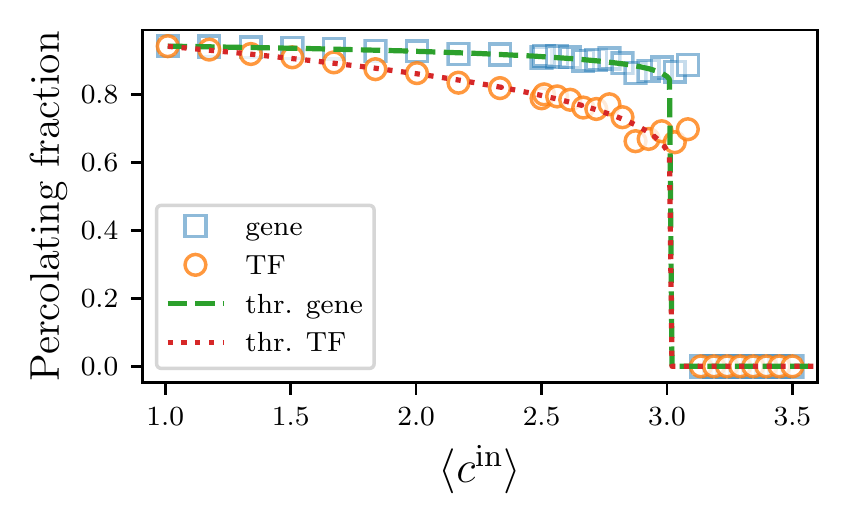}} \quad
    \subfloat[][]{\includegraphics[width=0.47\textwidth]{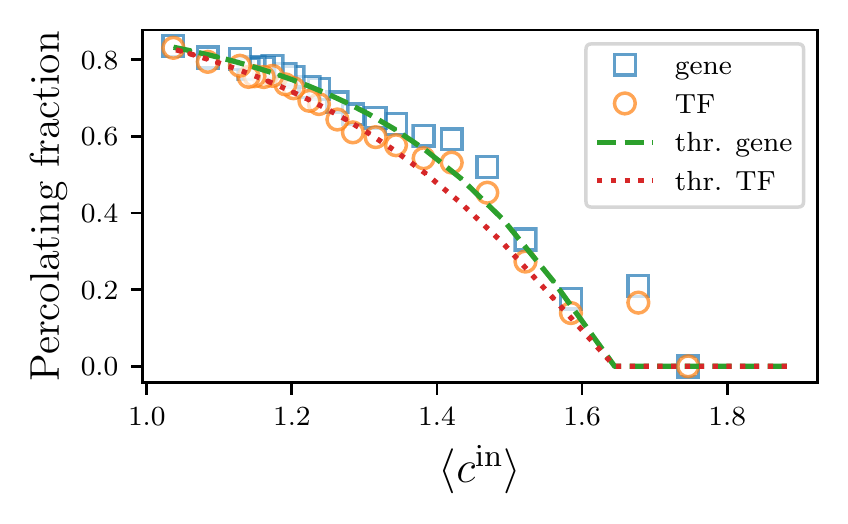}} \quad
\caption{\small Fraction of genes and TFs (squares and circle respectively) in aOC, consequent to removal of 5\% of genes, 
versus the average in-degree $\langle c^{\mathrm{in}}\rangle$ of TFs.
Results are shown for networks of {\bf (a)} type I with $\langle d^{\textrm{in}}\rangle=3$ and $N=10000$ and {\bf (b)} type II 
with $\langle d^{\textrm{in}}\rangle=1.4$ and $N=300000$.
Lines describe the theoretical prediction obtained from the largest solution of Eq.\,\eqref{eq:non-linear pert}. For the continuous transition shown in panel (b) the approach to the critical value $\langle c^{\rm in}\rangle_{\rm c}$ is linear $g\sim (\langle c^{\rm in}\rangle_{\rm c}-\langle c^{\rm in}\rangle)$.}
\label{fig:out_comp}
\end{figure}

We use simulations to validate the theoretical predictions above. We randomly generate  synthetic graphs with  prescribed connectivity, and determine the set of nodes that are in the aOC \textit{after} a random fraction $p$ of genes has been removed. We remove a TF if any of its predecessors is missing. We then re-evaluate the out-component after removal of any TFs and iterate this procedure until convergence. This pruning protocol ultimately gives the set of nodes in the giant aOC. 
We show results of this procedure, for the two families of networks of type I and type II, in Fig.\,\ref{fig:out_comp}. We fix the average 
value of the in-degree of genes $\langle d^{\mathrm{in}}\rangle$ and we tune the parameter $\langle c^{\mathrm{in}}\rangle$ in type I networks, and the exponent $\gamma$ of the fat tailed distribution in type II networks, for which $\langle c^{\textrm{in}}\rangle = \zeta(\gamma)$. Theoretical curves are given by the solution of  Eq.\,\eqref{eq:non-linear pert} with the largest values of $g^\star$ and $t^\star$. 
The fate of the giant aOC (if it exists) against deletion of genes is of interest for the interpretation of gene knock-out experiments in the limit where the number of genes deleted is actually finite. We will deal with this limit later on in Sect.\, \ref{sec:finite_p} below.

\subsection{Stability of non percolating phase }
\begin{figure}
    \centering
      \subfloat[][\label{graphical_nonpercolating}]{\includegraphics[width=0.47\textwidth]{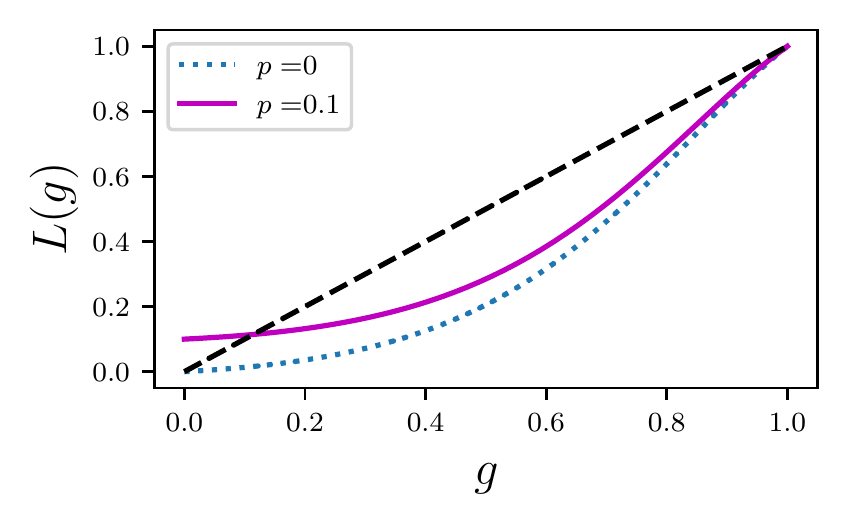}} \quad
          \subfloat[][]{\includegraphics[width=0.47\textwidth]{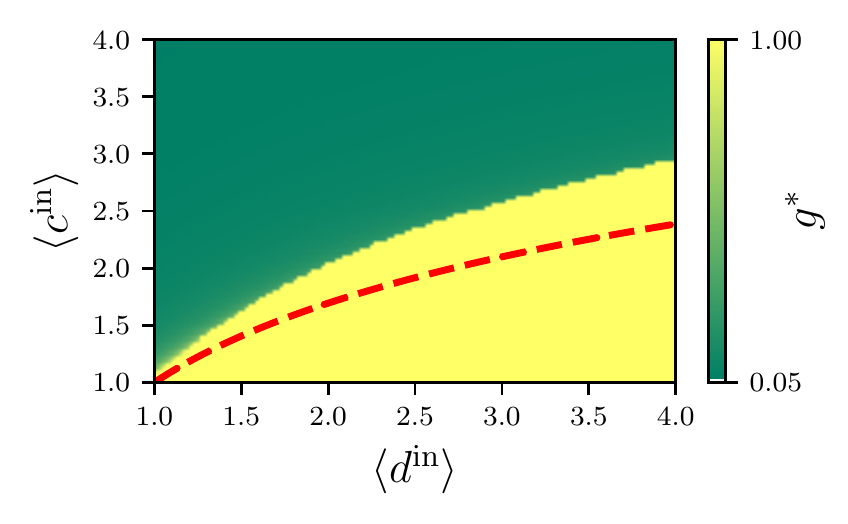}}  \\
    \caption{ \small \textbf{(a)} Graphical solution of  Eq.\,\eqref{eq:non-linear pert_off}. Comparison between the functional composition $L(g)$ describing the perturbed scenario at $p=0.1$ (solid line) and $F_1(g)$ describing the unperturbed scenario (dotted line). Results are shown for type I networks with connectivity $\langle c^{\textrm{in}}\rangle=$4, $\langle d^{\textrm{in}}\rangle=$2.  
    \textbf{(b)} Heatmap of 
    the largest solution $g^\star$ of Eq.\,\eqref{eq:non-linear pert_off} for a range of mean connectivities in type I networks. The fraction of genes in the giant aOC seed is $p=0.05$. The dashed line indicates the transition curve for $p=0$. 
    }
 \label{fig:non-percolating}
\end{figure}

We now investigate the stability of the solution $(g,t)=(0,0)$ using a perturbation of this solution which consists of finding a giant aOC which contains predefined randomly chosen set of genes that \textit{by definition} forms its seed (or nucleus). We consider a setting where the fraction $p$ of genes is thus defined to belong to the aOC, and ask under which conditions a giant aOC exists that contains this very set of genes. We assign a binary random variable $\chi_i$ taking the value $\chi_i=1$ if a gene is defined to belong to the seed of the aOC, and $\chi_i=0$ if it does not belong to that seed. We can analyse the existence of the giant aOC in terms of a set of indicator variables $n_i$ for genes and $\sigma_\mu$ for TFs as in Sect. \,\ref{sec:out} above. 
Following a line of reasoning as in the previous section, we obtain the modification of Eqs.\,\eqref{eq:cavity01} needed to capture the existence of a set of genes pre-defined to be the seed (or nucleus) of a giant aOC as 
\begin{equation}
\begin{aligned}
n_i&=&\left(1-\chi_i \right)\left(1-\prod_{\mu \in \partial _i^{\mathrm{in}}}(1-\sigma_\mu)\right)+\chi_i \ ,\\
\sigma_\mu&=&\prod_{i\in \partial _\mu^{\eta}}n_i\ ,
\end{aligned}
\label{eq:non-linear_perc_off}
\end{equation}
Averaging these over the realisations $\{\chi_i\}$  of the percolation (or seeding) experiment, we get
\begin{equation}
\begin{aligned}
g_i&=&\left(1-p \right)\left(1-\prod_{\mu \in \partial _i^{\xi}}(1-t_\mu)\right)+p \ ,\\
t_\mu&=&\prod_{i\in \partial _\mu^{\eta}}g_i\ ,
\end{aligned}
\label{eq:disorder_average_perc_off}
\end{equation}
for the probability of gene $i$ and TF $\mu$ to belong to a giant aOC thus seeded. Taking the average over the sites of a graph, as in Sect.\,\ref{sec:out}, one obtains
\begin{equation}
\begin{aligned}
g=&(1-p) \sum_{d=1}P^{\textrm{in}}_{\mathrm{D}}(d)\left[1-(1-t)^{d}\right]+p&=l_1(g)\ ,\\
t=& \sum_{c=1}P_{\rm C}^{\textrm{in}}(c)g^{c}&=l_2(t)\ .
\end{aligned}
\label{eq:non-linear pert_off}
\end{equation}
In the present case, $(g,t)=(1,1)$ is still always a solution, but $(g,t)=(0,0)$ is not for any $p \neq 0$. Depending on the degree distributions, there may now for $p \neq 0$ be two stable solutions: namely $(g,t)=(g^{\star},t^{\star})$ and $ (g,t)=(1,1)$. 
The functional composition 
$L(g)=l_1(l_2(g))$, that gives
the fraction of genes $g^{\star}$ solving \eqref{eq:non-linear pert_off} as $g^{\star}=L(g^{\star})$, can now be written in terms 
of the unperturbed functional composition $F_1(g)$, as $L(g)= F_1(g) +\left( 1-F_1(g) \right)p$. One has $L(g) \geq F_1(g)$ for 
$p\in (0,1)$. It is implied that when $(g,t)=(0,0)$ is the only stable solution in the unperturbed case, this will be continuously deformed when a perturbation is introduced, see Figure \ref{fig:non-percolating}a. Conversely, when the solution $(g,t)=(0,0)$ is unstable in the unperturbed case, a solution $(g,t)=(g^{\star},t^{\star})$ in the vicinity of $(0,0)$ \textit{will no longer exist} in the presence of perturbation. Hence, Eq.\,\eqref{eq:stability finite clusters} does indeed give the necessary condition for the stability of the non-percolating solution. 
In Fig.\,\ref{fig:non-percolating}b, the smallest solution $g^{\star}$ of Eq.\, \eqref{eq:non-linear pert}) is plotted as a function of 
the mean connectivities, for bipartite graphs of type I. Again, a discontinuous transition of $g^{\star}$, from a percolating to a non-percolating regime, is exhibited when connectivities are varied, 
as in Fig.\,\ref{fig:out non-linear analitic}. 
 
 \begin{figure}
 \centering
     \subfloat[][\label{on}]{\includegraphics[width=0.47\textwidth]{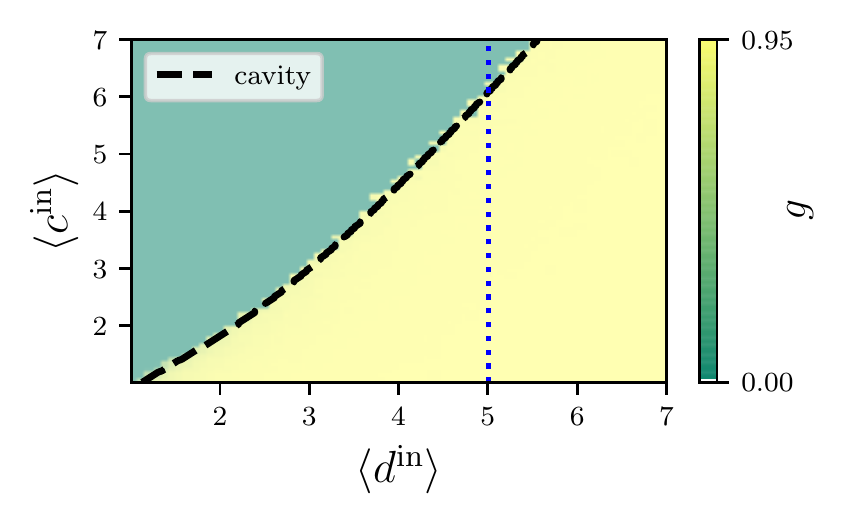}} 
    \subfloat[][\label{off}]{\includegraphics[width=0.47\textwidth]{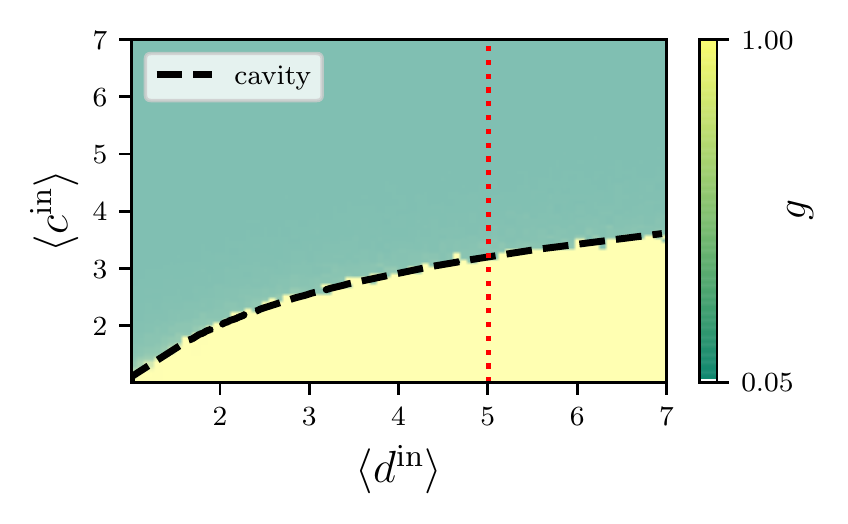}} \\
    \subfloat[][\label{histeresys}]{\includegraphics[width=0.47\textwidth]{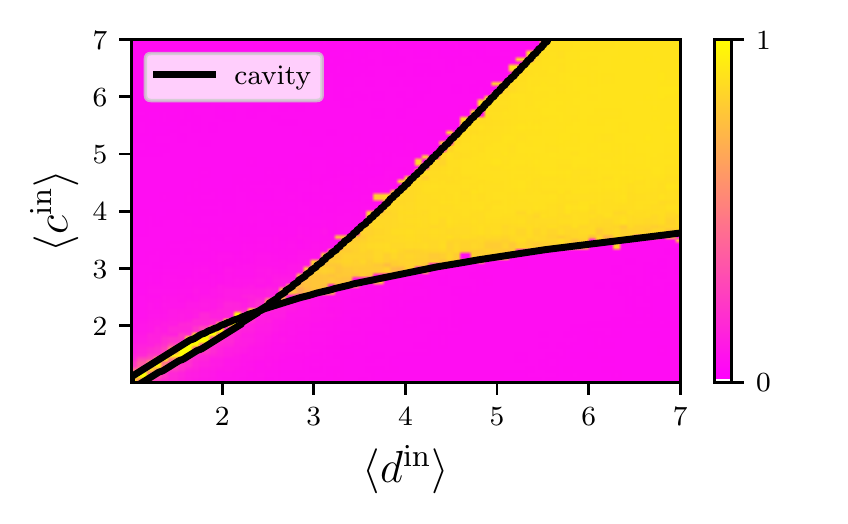}}
    \subfloat[][\label{slice}]{\includegraphics[width=0.47\textwidth]{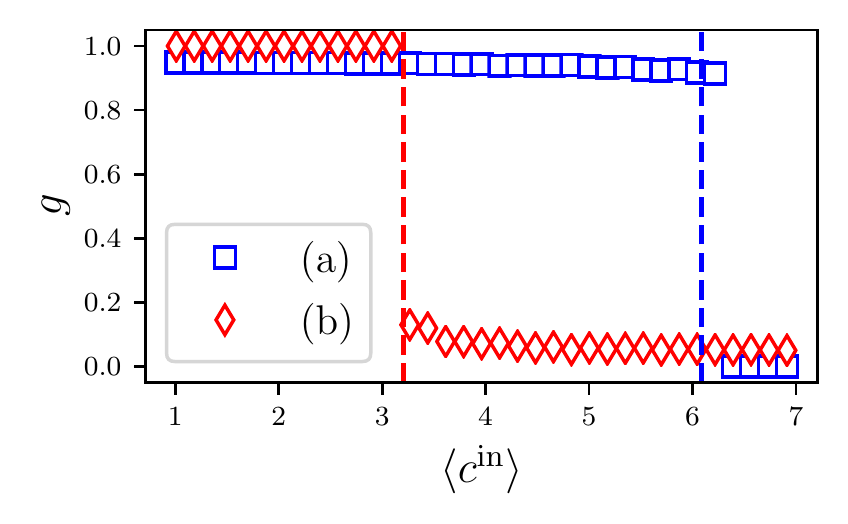}} 
\caption{\small Comparison between cavity predictions (Eqs. \eqref{eq:non-linear pert}, \eqref{eq:non-linear pert_off}) 
and the stationary state of the corresponding microscopic 
dynamics (Eqs. \eqref{eq:dynamics_non-linear_perc_dyn},  \eqref{eq:dynamics_non-linear_perc_off} respectively), 
averaged over the network sites. \textbf{(a)} Heat-map of the average gene expression level $g=N^{-1}\sum_i n_i$ for a configuration where a fraction $1-p$ of genes is clamped to be in inactive state, versus the mean connectivities of type I networks. \textbf{(b)} Heat-map of the average gene expression level for a configuration where a fraction $p$ of genes is clamped to be in active state. \textbf{(c)} Difference between the two protocols. \textbf{(d)} Plot of the average gene activation for a type I  
network with connectivity $\langle d^{\mathrm{in}}\rangle=5$, 
as indicated by the dotted lines in figures (a) and (b). The results 
of the protocols used in panels (a) and (b) are represented in terms of circles and stars, respectively. The dashed lines represent the cavity predictions for the instabilities. Simulation parameters are $N=10000$, $p=0.95$.}
\label{fig:non-linear}
\end{figure}

\subsection{Single instance cavity}
In order to study the stability of the percolating phase for a single instance of a GRN, we investigate the stability of 
the solutions of the microscopic Eqs.\,\eqref{eq:dynamics_non-linear_perc}. 
To this purpose, 
we consider 
the dynamical processes
\begin{equation}
\begin{aligned}
n_i(\tau+1)&=&\chi_i\left(1-\prod_{\mu \in \partial _i^{\mathrm{in}}}(1-\sigma_\mu(\tau))\right)\, \\
\sigma_\mu(\tau)&=&\prod_{i\in \partial _\mu^{\mathrm{in}}}n_i(\tau) \ .
\end{aligned}
\label{eq:dynamics_non-linear_perc_dyn}
\end{equation} 
whose stationary solutions  
give the stable solutions of Eqs.\,\eqref{eq:dynamics_non-linear_perc}, under forward iteration. 
Their average, over the network sites, is 
plotted in Fig.\,\ref{fig:non-linear}a 
and exhibits a discontinuous transition, as the network connectivity 
parameters are varied.  
Its location  
is correctly captured by the stability curves predicted by the macroscopic cavity analysis Eq.\eqref{eq:non-linear pert}.
The iterative dynamics of Eq.\, \eqref{eq:dynamics_non-linear_perc_dyn} is a special case of the gene regulatory dynamics of Eq.\,\eqref{eq: dynamics} with $\theta_i = 1-\chi_i$ and positive regulatory couplings. This version of the dynamics describes a setup in which a certain set of genes is \textit{forced} to remain silent, \textit{e.g.} as a result of a gene knockout experiment.

We now perform an analogous single instance analysis to the stability of the non-percolating phase by resorting to a solution of Eqs.\,\eqref{eq:non-linear_perc_off} through forward iteration:
\begin{equation}
\begin{aligned}
n_i(\tau+1)&=&\left(1-\chi_i \right)\left(1-\prod_{\mu \in \partial _i^{\mathrm{in}}}(1-\sigma_\mu(\tau)\right)+\chi_i \ ,\\
\sigma_\mu(\tau)&=&\prod_{i\in \partial _\mu^{\mathrm{in}}}n_i(\tau)\ .
\end{aligned}
\label{eq:dynamics_non-linear_perc_off}
\end{equation}
Figure \ref{fig:non-linear}b shows the site average of the stationary solution of the microscopic dynamics \eqref{eq:dynamics_non-linear_perc_off}. Just as for the percolating phase, the location of the transition in the stationary state of the  microscopic dynamics is correctly predicted by the cavity analysis  \eqref{eq:non-linear pert_off} carried out at the macroscopic level. 
As expected from a discontinuous transition, there is a region with coexisting solutions (Fig.\,\ref{fig:non-linear}c) which manifests itself as hysteresis in the dynamics: knocking out a small 
fraction of genes in a system where genes are initially on, will switch off the entire network, at a value of  connectivity which is larger than the one required for a small fraction of initially active genes 
(in a system of inactive genes)
to trigger an activation cascade
(Fig.\,\ref{fig:non-linear}d).

The iterative dynamics in Eqs. \eqref{eq:dynamics_non-linear_perc_off} is a special case of the gene regulatory dynamics in Eq.\,\eqref{eq: dynamics} with $\theta_i = -\chi_i$ and positive regulatory couplings. This dynamics describes a setup where a certain set of genes is clamped to be active, \textit{e.g.} as a result of an abundance of TFs 
passed from a mother cell to a daughter cell in the first moments after division.

It is worth noting that there is a second level of single instance cavity dynamics that arises from solving Eqs.\,\eqref{eq:cavity disorder average} and (\ref{eq:disorder_average_perc_off})
through forward iteration. Solutions agree with those of microscopic and macroscopic dynamics, unless the number of perturbed genes is $\mathcal{O}(1)$, creating finite sample effects which are particularly pronounced in the proximity of any phase transition. We discuss this behaviour in the next section.

\subsection{Small perturbation limit and analysis of gene knockout experiments}
\label{sec:finite_p}
In section \ref{sec: perturbation on} above we studied the stability of the giant aOC by removing a fraction $1-p$ of genes and computing the fraction of nodes that do not belong to giant aOC as a result of that removal. We refer to this quantity as the avalanche size. Here we investigate the avalanche size in the limit $p\rightarrow 1$, in particular with scaling $p = 1-\rho/N$. In this situation,  a finite number of genes are initially removed from the network, as opposed to the situation analysed in section \ref{sec: perturbation on}, where the number of genes that are initially removed from the network grows linearly with the network size $N$.

\begin{figure}
\centering
\includegraphics[width=0.6\textwidth]{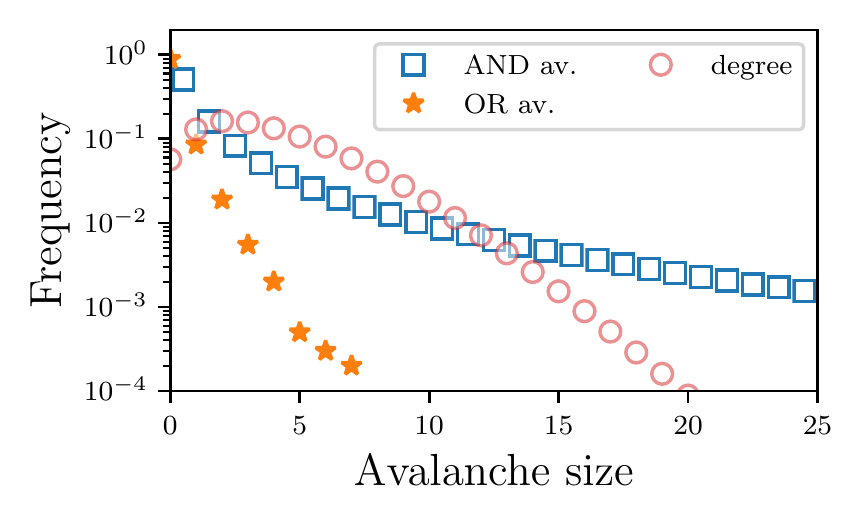}
\caption{Histogram of the avalanche size (only genes) \textit{i.e.} the number  of genes that are removed from the network as a result of a 
 single-gene knockout experiment. Sum of the heights is normalised to 1. The histogram for a single graph is computed. For each bin, the height is averaged over 5000 network realisations, and the average is shown (squares).  Error bars, computed as the standard deviation of the mean, are smaller than point markers.  Squares show results for the AND logic dynamics, while stars indicate results when OR gate logic is used in the same network. Circles indicate the out-degree distribution of the projected gene-to-gene graph $P^{\mathrm{out}}_{D,P}$. 
 Parameters: $\langle d^{\textrm{in}}  \rangle=$2 $\langle c^{\textrm{in}} \rangle=$2, $N=$10000.}
    \label{fig: stability out}
\end{figure}

This regime is relevant to the analysis of 
gene knockout experiments. These are performed in vitro as a tool to investigate the underlying (unknown) network. In our simulation, a synthetic network is generated and a single gene $i$ is removed. 
The dynamics of single instance cavity \eqref{eq:dynamics_non-linear_perc_dyn}, is then run 
with $\chi_j=1-\delta_{ij}$ 
and initial conditions $n_i(0)=1\ \forall i \in\lbrace 1,\dots N \rbrace $, until stationarity. 
In a realistic setting, we expect the GRN to consist of a robust giant aOC. We therefore consider networks that satisfy the stability condition of Eq.\,\eqref{eq:stability_out-comp}. For each gene in the original giant aOC, we compute the avalanche size associated with the removal of that gene. A histogram of avalanche sizes for networks of type I is shown in Fig.\,\ref{fig: stability out}. 

Here, we compare results for dynamics with 
different logic gates, the AND logic defined in Eqs.\, \eqref{eq: dynamicsg}, \eqref{eq: dynamics}
and an OR logic dynamics, obtained by replacing Eq.\, \eqref{eq: dynamics}) with 
\begin{equation}
\sigma_\mu(\tau)=\frac{1}{c_{\mu}^{\rm in}}\sum_{i\in \partial_\mu^{\rm in}} n_i(\tau)
\label{eq:OR}
\end{equation}
The distribution of avalanche sizes is very different for the two 
types of dynamics, emphasizing the crucial role of the underlying dynamics in interpreting data from gene knock-out experiments.
We also provide comparisons of these quantities with the out-degree distribution of the so-called ``projected graph" of effective 
gene-gene interactions,  defined by the adjacency matrix 
$A_{ij} = \Theta\big(\sum_{\mu}^{\alpha N}|r_{i\mu}| m_{\mu j}\big)$  \cite{hannam2019percolation}, where TFs have been integrated out (i.e. summed over). The out-degree distribution for this graph is defined by 
\begin{equation}
P^{\mathrm{out}}_{D,P} (k) =\frac{1}{N}\sum_{j}\left\langle \delta_{\sum_{i} A_{ij},k}\right\rangle \, \ ,
\label{eq: projected}
\end{equation}
in which the angled bracket denotes an average over the realisations of the membership and regulatory matrices $\mathbf{m}$ and $\mathbf{r}$, respectively. 
This fully characterises direct interactions in a network with the dynamics governed by the OR logic  Eq.\, \eqref{eq:OR} \cite{hannam2019percolation}.
However, the projected graph is not sufficient to fully 
specify the interactions in the case with the dynamics governed by the AND logic Eq.\,\eqref{eq: dynamics}, where it can only
give interaction pathways.

Unsurprisingly, the avalanche size distribution and the out-degree of the projected graph have different behaviour, even when the 
dynamics is governed by the OR logic. 
This is due to the fact that the avalanche process depends 
on network structure beyond the degree distribution. 
Hence, using the distribution of avalanche sizes in gene knock-out experiments, as proxies for the degree distribution of the gene-regulatory network, may systematically and significantly bias the tail of the distribution, in a way that depends on the underlying 
logic.

Let us now briefly also consider the small perturbation limit in a network for which the giant aOC is not robust, \textit{i.e.} for which the stability condition of  Eq.\,\eqref{eq:stability_out-comp} is violated. 
In Sect.\,\ref{sec: perturbation on} 
we found that Eq. \,\eqref{eq:stability_out-comp} gives the right 
stability criterion for the survival of the giant aOC, against 
removal of a {\it finite fraction} $1-p$ of nodes, for any $p$, and the perturbed cavity equations \eqref{eq:non-linear pert} converge, in the limit $p\rightarrow 1$, to the unperturbed Eqs.\,\eqref{eq: out non linear},
however, here, we are interested in the question of whether a \textit{single} node removal from a network will cause the 
fragmentation of the giant component,
in the regime where the latter is unstable. 
The outcome of such an experiment, 
will depend on the local topological properties of the node removed 
(see Ref. \cite{kuhn2017heterogeneous}), and cannot be directly 
extrapolated from the $p\rightarrow 1$ limit 
of the macroscopic equations derived in Sect.\,
\ref{sec: perturbation on}. 
If the number of nodes removed is small and does not grow with the size of the system, then the final state of the microscopic Eq.\,\eqref{eq:dynamics_non-linear_perc_dyn} will depend on the random  $\{\chi_i\}$ realisation. 
The effects of these fluctuations are particularly pronounced in the vicinity of the instability line, and will lead to a 
rounding of the phase transition 
over and above the one illustrated in Fig.\,\ref{fig:non-linear}a, where the number of removed nodes grew linearly with the size of the system. 
Results demonstrating this rounding from finite-size effects 
are presented and discussed in \ref{appendix}.

\section{Existence of a giant ``AND'' SCC}
\label{sec:SCC} 
 In this section we investigate the existence and stability of the ``AND'' strongly connected component (aSCC) using the cavity method, in a similar fashion as done above. We also present an algorithm that computes the giant aSCC for a given graph realisation, through a pruning procedure. We then verify the agreement between theory and simulations.  The starting point of our analysis is the observation that the set of nodes in a strongly connected component (SCC) is the intersection of the set of nodes in the in-component and out-component (OC) and this holds true when the ``AND" logic is introduced. We have 
 computed the fraction of nodes in the ``AND" out-component
 (aOC) in the previous section. To evaluate the fraction of nodes in 
 the ``AND" in-component, we exploit the fact that the ``AND'' condition is satisfied by construction in the in-component, as discussed in section \ref{sec:out}, hence the equations for the in-component, derived in the next section, also apply to the ``AND" in-component.

\subsection{Giant in-component}
To compute the fraction of nodes in the in-component, 
we introduce an indicator variable $n_i\in \lbrace 0,1\rbrace$ for each gene $i$, which is 1 if $i$ belongs to the giant \textit{in-component}, and $0$ otherwise. Analogously, let $\sigma_\mu\in\lbrace 0,1\rbrace$ denote the variable indicating whether or not the TF $\mu$ belongs to the giant in-component.
A node (either gene or TF) is in the giant \textit{in-component} if at least one of its \textit{successors} is in the giant \textit{in-component}, too; these conditions are captured by 
\begin{equation}
\begin{aligned}
n_i=1-\prod_{\nu \in \partial _i^{\mathrm{out}}}(1-\sigma_\nu^{(i)})\ ,\\
\sigma_\mu=1-\prod_{ j \in \partial _\mu^{\mathrm{out}}}(1-n_j^{(\mu)}) \ , \\
n_i^{(\mu)}=1-\prod_{\nu \in \partial _i^{\mathrm{out}}\setminus{\mu}}(1-\sigma_\nu^{(i)})\ ,\\
\sigma_\mu^{(i)}=1-\prod_{ j \in \partial _\mu^{\mathrm{out}}\setminus{i}}(1-n_j^{(\mu)}) \, \\
\end{aligned}
\label{eq:in-comp}
\end{equation}
where $\partial _i^{\mathrm{out}}$  denotes the set of TFs  $\nu$ which are successors  of gene  $i$. Thanks to  the almost sure equality of cavity and non-cavity indicator variables, as previously discussed  in Sect. \, \ref{sec:out_stability}, Eqs.\, \eqref{eq:in-comp} will produce a pair of self-consistency equations for the average fractions 
of nodes in the giant in-component 
which do not involve cavity variables,
\begin{equation}
\begin{aligned}
g=\sum_{d=1}P_D^{\textrm{out}}(d)\left[1-\left( 1-t\right)^{d}\right]
\ ,\\
t=\sum_{c=1}P_C^{\textrm{out}}(c)\left[ 1-\left(1-g\right)^{c}\right]\ .
\end{aligned}
\end{equation}

\subsection{Giant strongly connected component}
As explained above, for a node to belong to the strongly connected component (SCC), it must belong to the {\it intersection} of the in- and out- components, \textit{i.e.} gene $i$ is in the SCC  only if $i$'s  predecessors belong to the out-component and its successors belong to the in-component. 
To compute the fraction of nodes in the giant SCC, we 
introduce $\widehat{n}_i, \widecheck{n}_i, \bar{n}_i$ as the indicator variables for gene $i$ to belong to the giant in-, out-, and strongly connected components, respectively. Then $\bar{n}_i=\hat{n}_i, \widecheck{n}_i$, and the probability that a node belongs to the giant SCC is $\bar{g} = N^{-1} \sum_i \widecheck{n}_i \widehat{n}_i$. If distributions of in- and out- degreee of nodes are mutually independent, we simply get
\begin{eqnarray*}
\bar{g}=\widehat{g}\cdot \widecheck{g}\ ,\\
\bar{t}=\widehat{t}\cdot \widecheck{t}\ .
\end{eqnarray*}
where $\widehat{g}$   
and $\widecheck{g}$ 
are the probabilities of genes to be in the giant in- and out- components respectively, and $\widehat{t}$ and $\widecheck{t}$ denote 
the same quantities for TFs.

Here, we are interested in the existence of a giant strongly connected component, in the presence of the ``AND" logic, i.e. in the giant aSCC, rather than SCC. 
Since the in-component and the ``AND" in-component coincide, 
the difference between the giant SCC and aSCC resides in the 
difference between the giant OC and aOC. 
We recall that the solution of the macroscopic equations for the aOC describe situations where either the giant aOC is the entire network, 
or there is no giant aOC at all. In the former case, the aOC must also 
coincide with the OC, as the set of nodes in the OC must include 
those in the aOC. 
This means that the fraction of nodes belonging to the giant aSCC is either the same as that for the giant SCC, or zero (see Fig.\,\ref{fig: non-linear SCC}). 
In the next section, we will have 
a closer look at the topological properties of the aSCC.

\subsection{Algorithm to identify the giant aSCC}

The giant SCC of a network can be computed by identifying all the sites that can be reached from individual nodes, using \textit{e.g.} Tarjan's algorithm \cite{tarjan1972depth}. However, in the presence of the 
``AND" logic, the collection of paths emanating from individual nodes 
is no longer sufficient to identify the SCC. To circumvent this difficulty, we identify the aSCC through the following pruning procedure. We first identify the giant SCC using Tarajan's algorithm and denote with $\mathcal{A}$ the set of nodes belonging to the original graph but not to the SCC subgraph. We then prune all the nodes in $\mathcal{A}$ from the original graph. This results in some TFs no longer satisfying the ``AND'' constraint. These are then also removed and these steps are repeated until no new nodes are removed.

\begin{figure}
\centering
\includegraphics[width=0.8\textwidth]{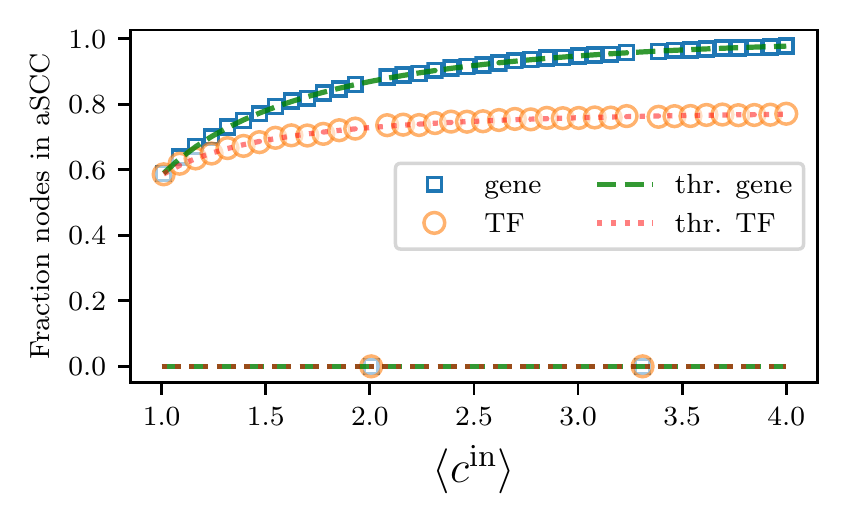}
\caption{\small Fraction of genes and TFs (squares and circle respectively) in aSCC versus the mean in-degree of TFs $\langle c^{\mathrm{in}}\rangle$. The average in-degree of genes is held constant $\langle d^{\mathrm{in}}\rangle = 1.5$. Points are obtained from simulations for a single network with size $N=50000$. Continuous lines represent the theoretical prediction for the fraction of nodes in SCC. 
}
\label{fig: non-linear SCC}
\end{figure}

We have shown in Fig.\,\ref{fig: non-linear SCC} that the fraction of nodes belonging to the giant aSCC is either 
the same as the fraction belonging to the giant SCC, or it is zero. We now discuss why this is the case from an algorithmic prospective. 
The (giant) SCC differs from the (giant)  aSCC if and only if there exists a gene which is 
not itself in the (giant) SCC and is a predecessor of at least one TF 
in the (giant) SCC; see Fig.\,\ref{fig:in-SCC-out}b for an example.
This would imply that none of the predecessors of that gene (called $k$ in  Fig.\,\ref{fig:in-SCC-out}b) %(which does not belong to the giant SCC)
belongs to the giant SCC either. 
Since regulatory constraints on the network imply that 
each genes has always at least one predecessor and at least one successor, by iterating this argument, 
one would exhaust the entire network, contradicting the initial hypothesis of the existence of a giant SCC, unless the set of predecessors form a loop. Therefore, a TF $\mu$ belonging to the (giant) SCC, will not satisfy the ``AND'' constraint if, and only if, its predecessors form a loop. In this case, TF $\mu$ does not belong to aOC. It is known that the probability of having a finite cluster scales as $N^{-1}$ with system size \cite{newman2001random}. 
Thus, the effect of removal of a finite number of nodes from the aOC is 
observed to lead to either the collapse of the giant aOC, or to the removal of a finite number of nodes, as shown in Fig.\,\ref{fig: stability single out}.\footnote{The probability that a node elimination triggers an avalanche is different from that computed in Fig.\,\ref{fig: stability single out}.  In figure \ref{fig: stability single out}a nodes are sampled uniformly from the network.} 
Therefore the pruning procedure starting from the SCC can produce the elimination of  (statistically)  all nodes in the network, or none.

\section{Summary and discussion}
\label{sec:conclusions}
We re-investigate the problem of percolation in a bipartite directed network model of gene regulation, that is based on the coupled dynamics of genes and transcription factors. The percolation problem plays a central role in assessing the viability of a GRN to sustain multi-cellular life under the realistic assumption that a GRN's dynamics is governed by noise. As shown in \cite{hannam2019percolation} the relevant percolation problem is that of heterogeneous $k$-core percolation on a directed bipartite graph. A re-investigation of the problem posed in \cite{hannam2019percolation} is needed, as fundamental constraints required for the degree distribution of GRNs were not properly taken into account in that investigation. Interestingly, we found that taking these constraints into account in the analysis leads to the most favourable situation regarding the size of the giant or percolating component, \textit{i.e.}, there is always a solution for which \textit{all} nodes of the network belong to the giant out component. The implication of this observation, however, depends very much on the stability of the various solutions found in the problem: investigating the  size of the  giant ``AND'' out-component  when  connectivity of the network is varied, we identify two families  of networks --- networks of type I displaying a discontinuous transition, and networks of type II exhibiting a  continuous  percolation transition. We found that networks constructed in terms of `shifted' Poisson degree distributions belonged to the type I class, whereas networks for which the out degrees of TFs follow a power-law distribution were found to be examples of type II networks.

We analyse the stability of the percolating and non-percolating phase using node percolation protocols. In particular, to assess the stability of the percolating phase, we study the robustness of a giant out-component against random node removal. Conversely, to assess the stability of the non-percolating phase, we find the smallest giant aOC which contains a predefined randomly chosen set of genes that by definition forms its seed.

We complement the cavity analysis at the global level by investigating iterative solutions of the microscopic cavity equations of single problem instances. This form of iterative solution can be interpreted as a particular realization of the gene expression dynamics described by Eqs.\,\eqref{eq: dynamicsg} and \eqref{eq: dynamics} in the presence of external signalling. The dynamical protocol used to study the  stability of the percolating phase mimics the genome-wide effect 
of an experiment that silences 
a set of genes (as in gene knock-out experiments). In the case of the stability of the non-percolating phase, the dynamical protocol mimics the activation of a GRN from a small set of active genes set externally to be active. This aims to sketch a potential mechanism for activation of all regulatory machinery of  the daughter cell from the transcriptome of the mother cell.

In the configuration where a stable aOC exists, the effect of single node percolation produces a cascade of node removals from the giant aOC. However, our simulation shows that this cascade is limited in size. Some experimental works on gene knockout \cite{kemmeren2014large} use the set of genes affected by single-gene knockout to infer a gene-gene interaction network. In the present paper, we highlight the fact that the network inferred based on knockout experiments is conceptually different from the network of gene interactions (if one were ever able to infer it with some confidence). In particular, the outcome of single-gene removal is an \textit{avalanche} of node removals (corresponding to gene-silencings) with a size distribution shown in Fig.\,\ref{fig: stability out}d. That size distribution does clearly not coincide with  the degree distribution of the underlying network. It is worth adding that the dynamics adopted in the present system involves only \textit{up}-regulating interactions, and that the statistics of avalanche sizes in systems with a mixture of up-regulating and down-regulating TFs will be different. Thus our results at this point merely serve as a warning that any inference of GRNs on the basis of results of gene knock-out experiments should be treated with sufficient levels of caution.

One of the results of our stability analyses is a minimal condition for the  full network to form a stable giant out component. The condition formulated in Eq.\, \eqref{eq:stability_out-comp} requires that $\langle c^\mathrm{{in}}\rangle P_{{D}}^{\mathrm{in}}(d=1)<1$.  This condition suggests that GRNs are characterised by a redundancy of TF binding affinities, \textit{i.e.} the fraction of genes that are regulated by only a single TF must be sufficiently small. Note, however, that the condition  Eq.\,\eqref{eq:stability_out-comp}  does not involve the actual (or average) number of TFs regulating a gene, 
and is still compatible with the assumption of TFs selectively binding to locations on the genome. The condition does, however, stipulate that the number of components forming  a single TFs complex must be sufficiently \textit{small on average}. Experimental evidences on known regulating complexes support the idea that the number of proteins constituting a TF is indeed generally small \cite{kribelbauer2019low}. We expect the condition of Eq.\,\eqref{eq:stability_out-comp} to be satisfied for \textit{any} gene regulatory network.  

The condition for the existence of a stable giant component represents a minimum requirement for the  existence of stable attractors of the dynamics of a GRN. In the presence of frustration (created through a combination of up-regulating and down-regulating couplings), the condition for the existence of a multiplicity of stable phases under noisy conditions is expected to be more stringent. To address this problem, we plan to study the (stochastic) dynamics in systems which include the effect of negative couplings.

An open challenge of our model 
is its calibration 
with available data.  Several databases, such as JASPAR \cite{sandelin2004jaspar} or TRANSFAC \cite{wingender2008transfac}, provide information about potential binding sites of TFs on the DNA that are based on motif-matching of binding domains on TFs and sites on the genome.  However, it is observed that motif-matching has a fairly low specificity for predicting TF binding sites \cite{marbach2016tissue}. To compensate for this, observed gene activity is often taken into account to increase the specificity of link prediction.  Our model is based on a clear distinction between gene activation states and the topology of the underlying network. Hence, to specify the gene expression dynamics, it is imperative to combine information related to gene dynamics with information about purely topological properties of the GRN. More specifically in this context, it should be noted that TF motif-matching does by itself not inform whether a TF is enhancing or inhibiting the expression of a given gene, let alone giving information about the strength of the regulatory effect. 
Information about the composition of  protein complexes can be extracted from databases such as CORUM \cite{ruepp2010corum} or the review \cite{morgunova2017structural}. However, the binding preferences of TFs that are protein complexes are not well known, which represents another challenge for the calibration of the model, and it  is currently unclear how to design an  experimental protocol   that would allow  providing the necessary information.

Elucidating the mechanisms of TF cooperation in gene regulation is an active field of research; we envisage that progress in the field can potentially give more comprehensive information to facilitate sensible network calibration in the near future.
 
 Finally,  while we have explored  gene regulation using the framework of   autonomous dynamical systems, future directions may include the role of external signalling in gene regulation. Moreover, it is worth emphasising that in the present paper we consider a model of gene-regulation entirely based on protein-coding genes. Yet, it is estimated that protein-coding genes only make up about 2\% of the entire genome \cite{di2018human}. It is believed that the non-coding genome, too, is involved in regulatory mechanisms at different stages of protein synthesis \cite{rinn2012genome}. 
An investigation of their role will constitute an interesting  pathway to future work. We are confident that the formulation of our model is flexible enough to allow us including general transcript-mediated regulation mechanisms in addition to protein-mediated regulation and effects of external signalling. 
The problems around calibrations mentioned above are, however, only going to be exacerbated in any such attempt.
 \section*{Acknowledgments}
GT supported by the EPSRC Centre for Doctoral Training in Cross-Disciplinary Approaches to Non-Equilibrium Systems (CANES EP/L015854/1). All authors thank Franca Fraternali and Joseph CF Ng for illuminating discussions. 
\section*{References}
\providecommand{\newblock}{}

%\bibliographystyle{iopart-num}
%\bibliography{iopart-num}

\newpage
\noindent
\appendix
\section{Single gene removal}
\label{appendix}

In this appendix we investigate the avalanche size resulting from removing a single gene in a type I 
network that does not have a robust giant aOC. 
\begin{figure}[h]
    \centering
    \subfloat[][]{\includegraphics[width=0.47\textwidth]{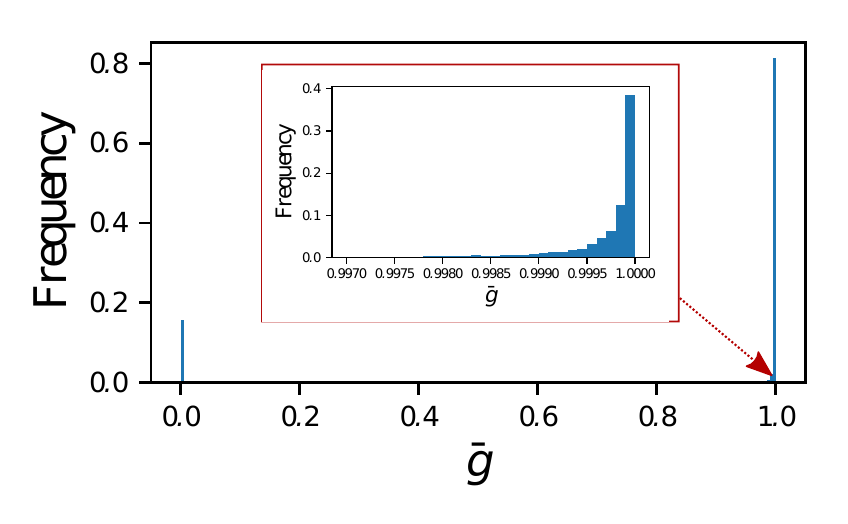}}
      \subfloat[][]{\includegraphics[width=0.47\textwidth]{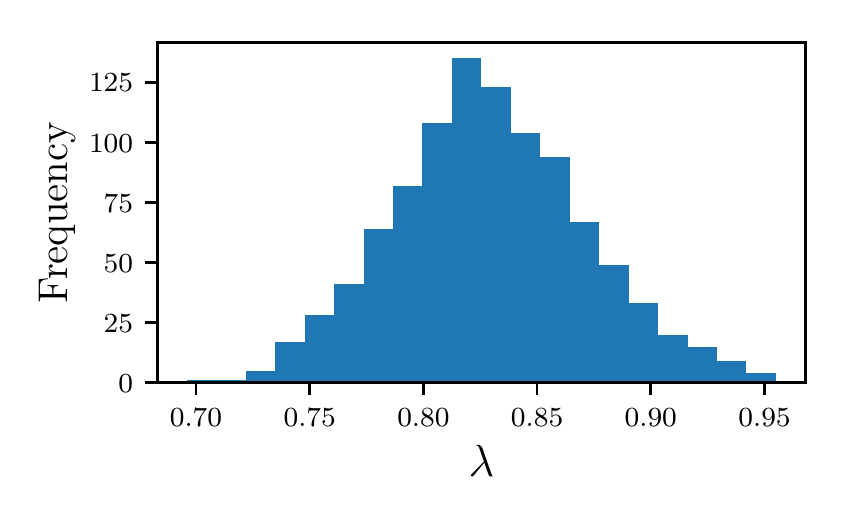}}      
      
    \subfloat[][]{\includegraphics[width=0.47\textwidth]{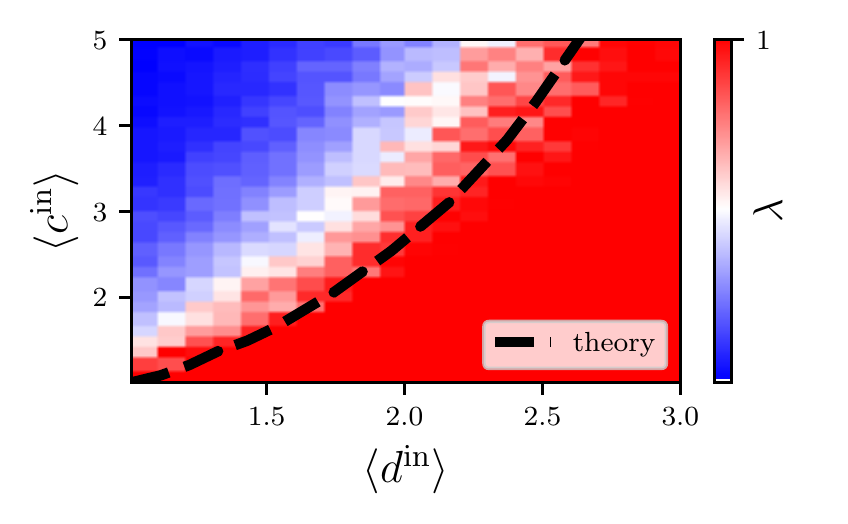}}
    \caption{\small %Stability of the aOC. 
   Effects of single-gene knock-out experiments on the aOC.
    \textbf{(a)}  Histogram of the fraction $\bar{g}$ of genes 
    remaining in the aOC as a result of 
    a single-gene knock-out experiment in a type I network with 
    $\langle d^{\textrm{in}} \rangle=2$ and $\langle c^{\textrm{in}} \rangle=$3.
    The values on the y-axis are re-scaled by the total number of knockout experiments, that is $N$. The inset magnifies the region $\bar{g}\simeq 1$.  \textbf{(b)} Histogram of the fraction 
    $\lambda$ (see definition in Eq.\,\eqref{eq: definition lambda}) of knockout experiments that do not trigger an extensive avalanche. 
    The histogram is computed over 1000 random network realisations 
    with the same connectivity as in (a). For each graph one value of $\lambda$ is evaluated. 
    \textbf{(c)}  Heat-map of $\lambda$ versus the mean degree of the two network layers. Each pixel represents a  graph. The dashed line represents the boundary of the inequality Eq.\,  \eqref{eq:stability_out-comp}.}
     \label{fig: stability single out}
    \end{figure}
Our analysis shows that, 
even in the 
range of connectivities where the macroscopic cavity analysis \eqref{eq:stability_out-comp} predicts the aOC to be unstable, the microscopic single instance dynamics \eqref{eq:dynamics_non-linear_perc_dyn} shows that a giant aOC 
may still be resilient to random removal of a \textit{finite} 
number of genes, at least for connectivities not too far away from 
the instability line. 
Here we look at the limiting case, where only a \textit{single} gene is removed from the network. 
For a graph with $N$ genes, we performed $N$ elimination experiments. Each experiment consists in removing one gene, say gene $j$, and running the dynamics \eqref{eq:dynamics_non-linear_perc_dyn} 
until stationarity, for
$\chi_i=1-\delta_{ij}$, and initial conditions $n_i(0)=1\ \forall i \in\lbrace 1,\dots N\rbrace$. For each experiment we compute $\bar{g}= N^{-1} \sum_i^{N} n_i(\infty)$, where $n_i(\infty)$ is the stationary state of the microscopic dynamics \eqref{eq:dynamics_non-linear_perc_dyn} for node $i$. The results of 
such
experiments are shown in Fig.\,\ref{fig: stability single out}. Even in the region where the macroscopic cavity analysis predicts only the solution $\bar{g}=0$, simulations occasionally exhibit a solution at $\bar{g}\simeq 1$. The histogram of $\bar{g}$, resulting from the $N$ single node elimination experiments, 
is shown in Fig.\,\ref{fig: stability single out}a, and presents two peaks, around $0$ and $1$. 
Magnifying the histogram in the region $\bar{g}\simeq 1$ (see inset), 
a substructure is observed which indicates that some percolation experiments only lead to a node removal cascade of \textit{finite} 
size.
We 
count the 
number of outcomes corresponding to the two peaks, and compute the 
fraction of genes whose removal does \textit{not} lead to an extensive elimination avalanche
 \begin{equation}
 \lambda=\dfrac{{\rm count}[\bar{g}\simeq 1]}{{\rm count}[\bar{g}\simeq 0]+ {\rm count}[\bar{g}\simeq 1]} \ .
\label{eq: definition lambda}
\end{equation}
Fluctuations in the value of $\lambda$ for different network instances are displayed in Fig.\,\ref{fig: stability single out}b, 
which shows the distribution of $\lambda$ over different network realisations 
of a family of random graphs 
with the same degree distribution. The connectivity parameters are chosen here 
such that the giant aOC is unstable
but connectivities are still 
\textit{close} to the instability line.  
The dependence of $\lambda$ on the network connectivities 
is shown in Fig.\,\ref{fig: stability single out}c as a  
heat-map, where each pixel corresponds to a network 
with a given combination of degrees. Results can be rationalised by noting that the parameter $\lambda$ can be interpreted as a stability measure, given that $\lambda=0$ characterises the condition of an unstable giant aOC, and $\lambda=1$ is indicative of a stable giant 
aOC.  
Fig.\,\ref{fig: stability single out}c shows that, 
in contrast to the results predicted from Eq.\/\eqref{eq:non-linear pert} and presented in Fig.\,\ref{on}, valid when a small but \textit{extensive} perturbation of the giant aOC is applied, the parameter $\lambda$ does not exhibit a discontinuous transition 
between the regimes where the non-percolating solution and the percolating solution encompassing (almost) the entire network are stable, respectively. As anticipated in Sect. \, \ref{sec:finite_p}, 
this is not unexpected: it is well known that in finite systems there will \textit{always} be some finite-size rounding of phase transitions \cite{green1983phase}, with additional subtleties expected for discontinuous phase transitions in disordered systems \cite{stiefvater1996averaging}. 
The analysis presented here demonstrates that 
fluctuations and rounding effects are magnified in the case of non-extensive perturbations, where self-averaging mechanisms are not expected to occur in individual perturbation experiments. Results also show that the finite-size rounding of the discontinuous transition is not symmetric relative to the transition line. Nonetheless, the instability line predicted from the macroscopic analysis does \textit{still} approximately identify the region where the outcome of 
a single-gene knockout experiment will give rise to extensive 
avalanches of node eliminations.

\end{document}